\documentclass{vldb}
\usepackage{graphicx}
\usepackage{balance}  
\usepackage{subcaption}
\usepackage{tabularx}
\usepackage{array}
\usepackage{booktabs}
\usepackage{rotating}
\usepackage{cite}
\usepackage{url}
\usepackage{enumitem}
\usepackage{float}
\usepackage{placeins}

\newcommand{\sparagraph}[1]{\vspace{1mm}\noindent {\bf #1}}

\usepackage{xcolor}
\newcommand{\hl}[1]{#1}

\begin{document}


\title{Bao: Learning to Steer Query Optimizers}




%
%
%
%

\numberofauthors{1} 

\author{
  \alignauthor
  Ryan Marcus$^{12}$, 
  Parimarjan Negi$^{1}$, 
  Hongzi Mao$^{1}$,\\
  Nesime Tatbul$^{12}$,
  Mohammad Alizadeh$^{1}$, 
  Tim Kraska$^{1}$ \\
  \affaddr{$^1$MIT CSAIL \quad $^2$Intel Labs}
  \email{\{ryanmarcus, pnegi, hongzi, tatbul, alizadeh, kraska\}@csail.mit.edu}
}

\maketitle


\begin{abstract}
  Query optimization remains one of the most challenging problems in data management systems. Recent efforts to apply machine learning techniques to query optimization challenges have been promising, but have shown few practical gains due to substantive training overhead, inability to adapt to changes, and poor tail performance. Motivated by these difficulties and drawing upon a long history of research in multi-armed bandits, we introduce Bao (the \underline{Ba}ndit \underline{o}ptimizer). Bao takes advantage of the wisdom built into existing query optimizers by providing per-query optimization hints.
  Bao combines modern tree convolutional neural networks with Thompson sampling, a decades-old and well-studied reinforcement learning algorithm. As a result, Bao automatically learns from its mistakes and adapts to changes in query workloads, data, and schema. Experimentally, we demonstrate that Bao can quickly (an order of magnitude faster than previous approaches) learn strategies that improve end-to-end query execution performance, including tail latency. In cloud environments, we show that Bao can offer both reduced costs and better performance compared with a sophisticated commercial system.
\end{abstract}

\section{Introduction}

Query optimization is an important task for database management systems. Despite decades of study~\cite{systemr}, the most important elements of query optimization -- cardinality estimation and cost modeling -- have proven difficult to crack~\cite{qo_unsolved}.

Several works have applied machine learning techniques to these stubborn problems~\cite{leo, deep_card_est, deep_card_est2, qo_state_rep, rejoin, sanjay_wat, neo, learn_cost, skinnerdb}. While all of these new solutions demonstrate remarkable results, they suffer from fundamental limitations that prevent them from being integrated into a real-world DBMS. Most notably, these techniques (including those coming from authors of this paper) suffer from three main drawbacks:

\begin{enumerate}[parsep=0mm, leftmargin=0em,labelwidth=*,align=left]
\item{{\it Sample efficiency.} Most proposed machine learning techniques require an impractical amount of training data before they have a positive impact on query performance. For example, ML-powered cardinality estimators require gathering precise cardinalities from the underlying data, a prohibitively expensive operation in practice (this is why we wish to estimate cardinalities in the first place). Reinforcement learning techniques must process thousands of queries before outperforming traditional optimizers, which (when accounting for data collection and model training) can take on the order of days~\cite{neo, sanjay_wat}.}
\item{{\it Brittleness.} While performing expensive training operations once may already be impractical, changes in query workload, data, or schema can make matters worse. Learned cardinality estimators must be retrained when data changes, or risk becoming stale. Several proposed reinforcement learning techniques assume that both the data and the schema remain constant, and require complete retraining when this is not the case~\cite{sanjay_wat, rejoin, qo_state_rep, neo}.}
\item{{\it Tail catastrophe.} Recent work~ has shown that learning techniques can outperform traditional optimizers \emph{on average}, but often perform catastrophically (e.g., 100x regression in query performance) in the tail\cite{neo,rejoin,learned_card_eval}. This is especially true when training data is sparse. While some approaches offer statistical guarantees  of their dominance in the average case~\cite{skinnerdb}, such failures, even if rare, are unacceptable in many real world applications.}
\end{enumerate}

\begin{figure}
\centering
\includegraphics[width=\linewidth]{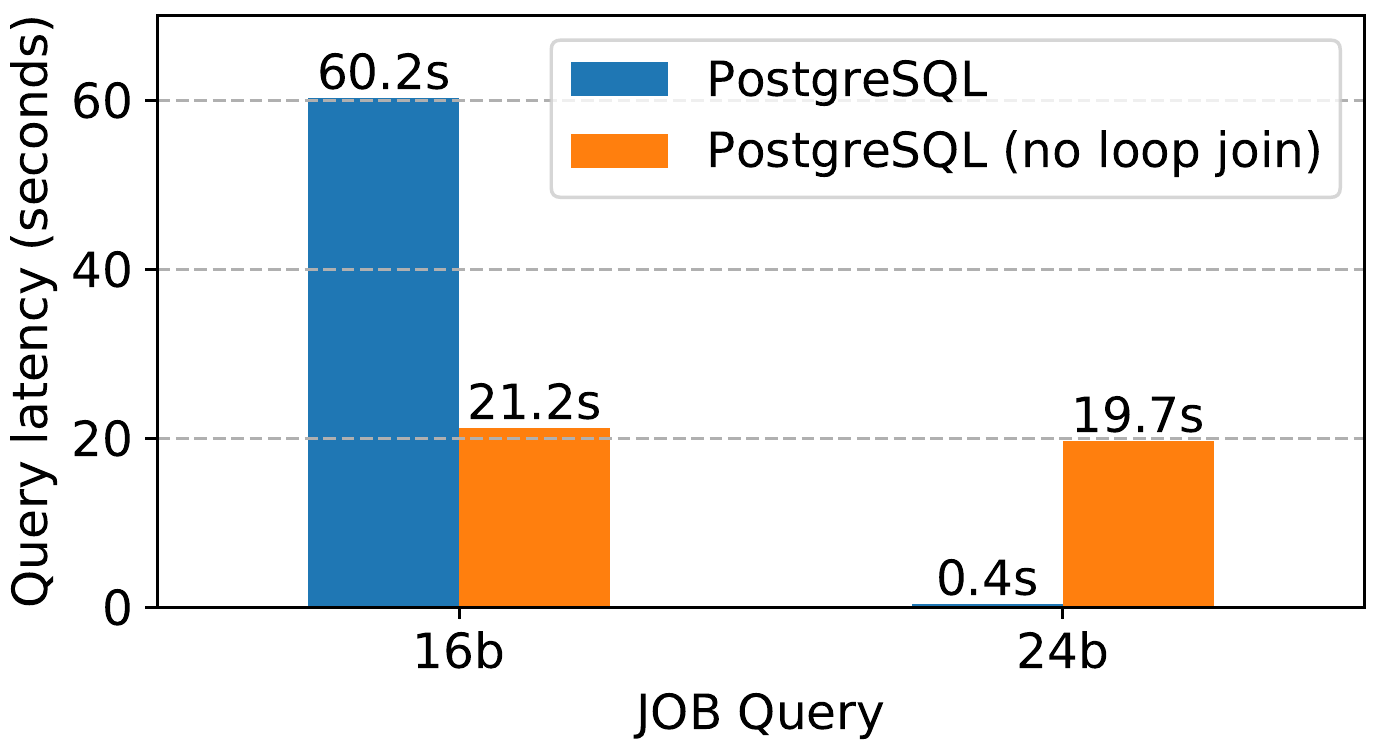}
\caption{Disabling the loop join operator in PostgreSQL can improve (16b) or harm (24b) a particular query's performance. These example queries are from the Join Order Benchmark (JOB)~\cite{howgood}.}
\label{fig:noloop}
\end{figure}

\sparagraph{Bao} Bao (\underline{Ba}ndit \underline{o}ptimizer), our prototype optimizer, can outperform traditional query optimizers, \hl{\emph{both open-source and commercial}}, with minimal training time ($\approx 1$ hour). Bao can maintain this advantage even in the presence of workload, data, and schema changes, all while rarely, if ever, incurring a catastrophic execution. While previous learned approaches either did not improve or did not evaluate tail performance, we show that Bao is capable of improving tail performance \emph{by orders of magnitude} after a few hours of training. Finally, we demonstrate that Bao is capable of reducing costs \emph{and} increasing performance on modern cloud platforms in realistic, warm-cache scenarios.


Our fundamental observation is that previous learned approaches to query optimization~\cite{neo, rejoin, skinnerdb, sanjay_wat}, which attempted to replace either the entire query optimizer or large portions of it with learned components, may have thrown the baby out with the bathwater. Instead of discarding traditional query optimizers in favor of a fully-learned approach, Bao recognizes that traditional query optimizers contain decades of meticulously hand-encoded wisdom. For a given query, Bao intends only to \emph{steer} a query optimizer in the right direction using coarse-grained hints. In other words, Bao seeks to build learned components on top of existing query optimizers in order to enhance query optimization, rather than replacing or discarding traditional query optimizers altogether.

For example, a common observation about PostgreSQL is that cardinality under-estimates frequently prompt the query optimizer to select loop joins when other methods (merge, hash) would be more effective~\cite{howgood, job2}. This occurs in query 16b of the Join Order Benchmark (JOB)~\cite{howgood}, as depicted in Figure~\ref{fig:noloop}. Disabling loop joins causes a 3x performance improvement for this query. However, for query 24b, disabling loop joins creates a serious regression of almost 50x. Thus, providing the coarse-grained hint ``disable loop joins'' helps some queries, but harms others.

At a high level, Bao sits on top of an existing query optimizer and tries to learn a mapping between incoming queries and the powerset of such hints. Given an incoming query, Bao selects a set of coarse-grained hints that limit the search space of the query optimizer (e.g., eliminating plans with loop joins from the search space). Bao learns to select different hints for different queries, discovering when the underlying query optimizer needs to be steered away from certain areas of the plan space.

Our approach assumes a finite set of query hints and treats each subset of hints as an arm in a contextual multi-armed bandit problem. While in this work we use query hints that remove entire operator types from the plan space (e.g., no hash joins), in practice there is no restriction that these hints are so broad: one could use much more specific hints. Our system learns a model that predicts which set of hints will lead to good performance for a particular query. When a query arrives, our system selects hints, executes the resulting query plan, and observes a reward. Over time, our system refines its model to more accurately predict which hints will most benefit an incoming query. For example, for a highly selective query, our system can steer an optimizer towards a left-deep loop join plan (by restricting the optimizer from using hash or merge joins), and to disable loop joins for less selective queries. This learning is automatic.

By formulating the problem as a contextual multi-armed bandit, Bao can take advantage of sample efficient and well-studied algorithms~\cite{thompson_intro}. Because Bao takes advantage of an underlying query optimizer, Bao has cost and cardinality estimates available, allowing Bao to use a more flexible representation that can adapt to new data and schema changes just as well as the underlying query optimizer. Finally, while other learned query optimization methods have to relearn what traditional query optimizers already know, Bao can immediately start learning to improve the underlying optimizer, and is able to reduce tail latency \emph{even compared to traditional query optimizers.}

Interestingly, it is easy to integrate information about the DBMS cache into our approach. Doing so allows Bao to use information about what is held in memory when choosing between different hints. This is a desirable feature because reading data from in-memory cache is significantly faster than reading information off of disk, and it is possible that the best plan for a query changes based on what is cached. While integrating such a feature into a traditional cost-based optimizer may require significant engineering and hand-tuning, making Bao cache-aware is as simple as surfacing a description of the cache state.

A major concern for optimizer designers is the ability to debug and explain decisions, which is itself a subject of significant research~\cite{develops, coko, opt_viz, opt_prov}. Black-box deep learning approaches make this difficult, although progress is being made~\cite{xai}. Compared to other learned query optimization techniques, Bao makes debugging easier. When a query misbehaves, an engineer can examine the query hint chosen by Bao. If the underlying optimizer is functioning correctly, but Bao made a poor decision, exception rules can be written to exclude the selected query hint. While we never needed to implement any such exception rules in our experimental study, Bao's architecture makes such exception rules significantly easier to implement than for other black-box learning models.


Bao's architecture is extensible. Query hints can be added or removed over time. Assuming additional hints do not lead to query plans containing entirely new physical operators, adding a new query hint requires little additional training time. This potentially enables quickly testing of new query optimizers: a developer can introduce a hint that causes the DBMS to use a different optimizer, and Bao will automatically learn which queries perform well with the new optimizer (and which queries perform poorly).

In short, Bao combines a tree convolution model~\cite{tree_conv}, an intuitive neural network operator that can recognize important patterns in query plan trees~\cite{neo}, with Thompson sampling~\cite{thompson}, a technique for solving contextual multi-armed bandit problems. This unique combination allows Bao to explore and exploit knowledge quickly. 

The contributions of this paper are:
\begin{itemize}
\item{We introduce Bao, a learned system for query optimization that is capable of learning how to apply query hints on a case-by-case basis.}
\item{We introduce a simple predictive model and featurization scheme that is independent of the workload, data, and schema.}
\item{For the first time, we demonstrate a learned query optimization system that outperforms both open source and commercial systems in cost and latency, all while adapting to changes in workload, data, and schema.}
\end{itemize}

The rest of this paper is organized as follows. In Section~\ref{sec:sysmod}, we introduce the Bao system model and give a high-level overview of the learning approach. In Section~\ref{sec:bandit}, we formally specify Bao's optimization goal, and describe the predictive model and training loop used. We present related works in Section~\ref{sec:rw}, experimental analysis in Section~\ref{sec:expr}, and concluding remarks in Section~\ref{sec:conclusion}.


\section{System model}
\label{sec:sysmod}

\begin{figure}
  \includegraphics[width=0.48\textwidth]{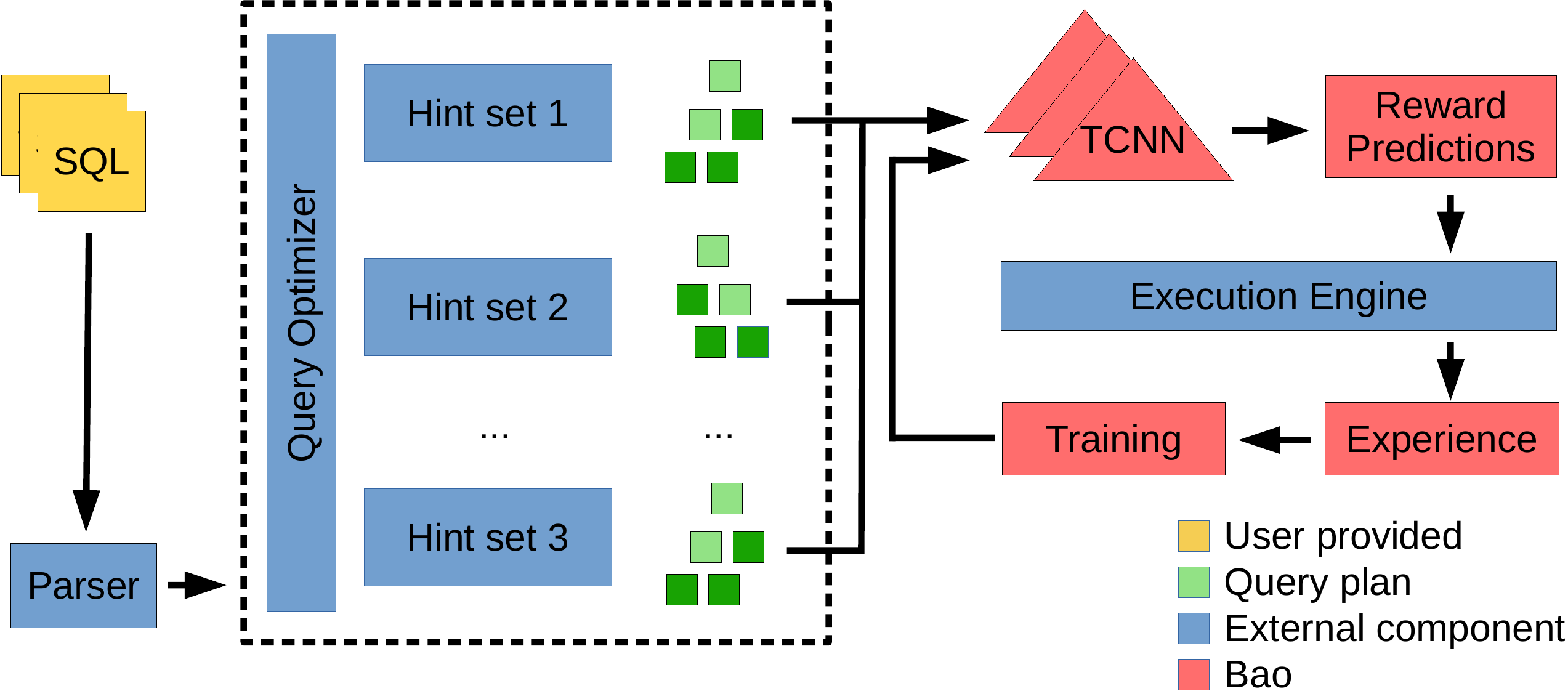}
  \caption{Bao system model}
  \label{fig:bao}
\end{figure}

Bao's system model is shown in Figure~\ref{fig:bao}. When a user submits a query, Bao's goal is to select a set of query hints that will give the best performance for the user's specific query (i.e., Bao chooses different query hints for different queries). To do so, Bao uses the underlying query optimizer to produce a set of query plans (one for each set of hints), and each query plan is transformed into a vector tree (a tree where each node is a feature vector).
These vector trees are fed into Bao's predictive model, a tree convolutional neural network~\cite{tree_conv}, which predicts the outcome of executing each plan (e.g., predicts the wall clock time of each query plan).\footnote{\hl{This process can be executed in parallel to maintain reasonable optimization times, see Section~\ref{sec:warm}.}}

Bao chooses which plan to execute using a technique called Thompson sampling~\cite{thompson} (see Section~\ref{sec:bandit}) to balance the exploration of new plans with the exploitation of plans known to be fast. The plan selected by Bao is sent to a query execution engine. Once the query execution is complete, the combination of the selected query plan and the observed performance is added to Bao's experience. Periodically, this experience is used to retrain the predictive model, creating a feedback loop. As a result, Bao's predictive model improves, and Bao learns to select better and better query hints.


\sparagraph{Query hints \& hint sets} Bao requires a set of \emph{query optimizer hints}, which we refer to as query hints or hints. A hint is generally a flag passed to the query optimizer that alters the behavior of the optimizer in some way. For example, PostgreSQL~\cite{url-pg_hints}, MySQL~\cite{url-mysql_hints}, and SQL Server~\cite{url-mssql_hints} all provide a wide range of such hints. While some hints can be applied to a single relation or predicate, Bao focuses only on query hints that are a boolean flag (e.g., disable loop join, force index usage). For each incoming query, Bao selects a \emph{hint set}, a valid set of query hints, to pass to the optimizer. We assume that the validity of a hint set is known ahead of time (e.g., for PostgreSQL, one cannot disable loop joins, merge joins, and hash joins at once). We assume that all valid hint sets cause the optimizer to produce a semantically-valid query plan (i.e., a query plan that produces the correct result).

\sparagraph{Model training} Bao's predictive model is responsible for estimating the quality of query plans produced by a hint set. Learning to predict plan quality requires balancing exploration and exploitation: Bao must decide when to explore new plans that might lead to improvements, and when to exploit existing knowledge and select plans similar to fast plans seen in the past. We formulate the problem of choosing between query plans as a \emph{contextual multi-armed bandit}~\cite{bandit_survey} problem: each hint set represents an arm, and the query plans produced by the optimizer (when given each hint set) represent the contextual information. To solve the bandit problem, we use Thompson sampling~\cite{thompson}, an algorithm with both theoretical bounds~\cite{thompson_infotheory} and real-world success~\cite{thompson_intro}. The details of our approach are presented in Section~\ref{sec:bandit}.

\sparagraph{Mixing query plans} Bao selects an query plan produced by a single hint set. Bao does \emph{not} attempt to ``stitch together''~\cite{stitch} query plans from different hint sets. While possible, this would increase the Bao's action space (the number of choices Bao has for each query). Letting $k$ be the number of hint sets and $n$ be the number of relations in a query, by selecting only a single hint set Bao has $O(k)$ choices per query. If Bao stitched together query plans, the size of the action space would be $O(k \times 2^n)$ ($k$ different ways to join each subset of $n$ relations, in the case of a fully connected query graph). This is a larger action space than used in previous reinforcement learning for query optimization works~\cite{rejoin, neo}. Since the size of the action space is an important factor for determining the convergence time of reinforcement learning algorithms~\cite{large_discrete}, we opted for the smaller action space in hopes of achieving quick convergence.\footnote{We experimentally tested the plan stitching approach using Bao's architecture, but we were unable to get the model to convergence.}


\section{Selecting query hints}
\label{sec:bandit}


Here, we discuss Bao's learning approach. We first define Bao's optimization goal, and formalize it as a contextual multi-armed bandit problem. Then, we apply Thompson sampling, a classical technique used to solve such problems. 

Bao assumes that each hint set $HSet_i \in F$ in the family of hint sets $F$ is a function mapping a query $q \in Q$ to a query plan tree $t \in T$:

\begin{equation*}
HSet_i : Q \to T
\end{equation*}

\noindent This function is realized by passing the query $Q$ and the selected hint $HSet_i$ to the underlying query optimizer. We refer to $HSet_i$ as this function for convenience. We assume that each query plan tree $t \in T$ is composed of an arbitrary number of operators drawn from a known finite set (i.e., that the trees may be arbitrarily large but all of the distinct operator \emph{types} are known ahead of time).

Bao also assumes a user-defined performance metric $P$, which determines the quality of a query plan by executing it. For example, $P$ may measure the execution time of a query plan, or may measure the number of disk operations performed by the plan.

For a query $q$, Bao must select a hint set to use. We call this selection function $B: Q \to F$. Bao's goal is to select the best query plan (in terms of the performance metric $P$) produced by a hint set. We formalize the goal as a regret minimization problem, where the regret for a query $q$, $R_q$, is defined as the difference between the performance of the plan produced with the hint set selected by Bao and the performance of the plan produced with the ideal hint set:

\begin{equation}
  \label{eq:regret}
  R_q = \left( P(B(q)(q)) - \min_i P(HSet_i(q)) \right)^2
\end{equation}

\sparagraph{Contextual multi-armed bandits (CMABs)} The regret minimization problem in Equation~\ref{eq:regret} can be thought of in terms of a contextual multi-armed bandit~\cite{bandit_survey}, a classic concept from reinforcement learning. A CMAB is problem formulation in which an agent must maximize their reward by repeatedly selecting from a fixed number of \emph{arms}. The agent first receives some contextual information (\emph{context}), and must then select an arm. Each time an arm is selected, the agent receives a \emph{payout}. The payout of each arm is assumed to be independent given the contextual information. After receiving the payout, the agent receives a new context and must select another arm. Each trial is considered independent. The agent can maximize their payouts by minimizing their regret: the closer the agent's actions are to optimal, the closer to the maximum possible payout the agent gets.

For Bao, each ``arm'' is a hint set, and the ``context'' is the set of query plans produced by the underlying optimizer given each hint set. Thus, our agent observes the query plans produced from each hint set, chooses one of those plans, and receives a reward based on the resulting performance. Over time, our agent needs to improve its selection and get closer to choosing optimally (i.e., minimize regret). Doing so involves balancing exploration and exploitation: our agent must not always select a query plan randomly (as this would not help to improve performance), nor must our agent blindly use the first query plan it encounters with good performance (as this may leave significant improvements on the table).

\sparagraph{Thompson sampling} A classic algorithm for solving CMAB regret minimization problems while balancing exploration and exploitation is Thompson sampling~\cite{thompson}. Intuitively, Thompson sampling works by slowly building up \emph{experience} (i.e., past observations of performance and query plan tree pairs). Periodically, that experience is used to construct a predictive model to estimate the performance of a query plan. This predictive model is used to select hint sets by choosing the hint set that results in the plan with the best predicted performance.

Formally, Bao uses a predictive model $M_\theta$, with model parameters (weights) $\theta$, which maps query plan trees to estimated performance, in order to select a hint set. Once a query plan is selected, the plan is executed, and the resulting pair of a query plan tree and the observed performance metric, $(t_i, P(t_i))$, is added to Bao's experience $E$. Whenever new information is added to $E$, Bao updates the predictive model $M_\theta$.  

In Thompson sampling, this predictive model is trained differently than a standard machine learning model. Most machine learning algorithms train models by searching for a set of parameters that are most likely to explain the training data. In this sense, the quality of a particular set of model parameters $\theta$ is measured by $P(\theta \mid E)$: the higher the likelihood of your model parameters given the training data, the better the fit. Thus, the most likely model parameters can be expressed as the expectation (modal parameters) of this distribution, which we write as $\mathbb{E}[P(\theta \mid E)]$. However, in order to balance exploitation and exploration, we \emph{sample} model parameters from the distribution $P(\theta \mid E)$, whereas most machine learning techniques are designed to find the most likely model given the training data, $\mathbb{E}[P(\theta \mid E)]$.

Intuitively, if one wished to maximize exploration, one would choose $\theta$ entirely at random. If one wished to maximize exploitation, one would choose the modal $\theta$ (i.e., $\mathbb{E}[P(\theta \mid E)]$). Sampling from $P(\theta \mid E)$ strikes a balance between these two goals~\cite{thompson_bound}. To reiterate, sampling from $P(\theta \mid E)$ is not the same as training a model over $E$. We discuss the differences at the end of Section~\ref{sec:tcnn}.

It is worth noting that selecting a hint set for an incoming query is not exactly a bandit problem. This is because the choice of a hint set, and thus a query plan, will affect the cache state when the next query arrives, and thus every decision is not entirely independent. For example, choosing a plan with index scans will result in an index being cached, whereas choosing a plan with only sequential scans may result in more of the base relation being cached. However, in OLAP environments queries frequently read large amounts of data, so the effect of a single query plan on the cache tends to be short lived. Regardless, there is substantial experimental evidence suggesting that Thompson sampling is still a suitable algorithm in these scenarios~\cite{thompson_intro}.

We next explain Bao's predictive model, a tree convolutional neural network. Then, in Section~\ref{sec:train_loop}, we discuss how Bao effectively applies its predictive model and Thompson sampling to query optimization.

\subsection{Predictive model}
\label{sec:model}

The core of Thompson sampling, Bao's algorithm for selecting hint sets on a per-query basis, is a predictive model that, in our context, estimates the performance of a particular query plan. Based on their success in~\cite{neo}, Bao use a tree convolutional neural network (TCNN) as its predictive model.
In this section, we describe (1) how query plan trees are transformed into trees of vectors, suitable as input to a TCNN, (2) the TCNN architecture, and (3) how the TCNN can be integrated into a Thompson sampling regime (i.e., how to sample model parameters from $P(\theta \mid E)$ as discussed in Section~\ref{sec:bandit}).

\subsubsection{Vectorizing query plan trees}
\label{sec:vectorize}
Bao transforms query plan trees into trees of vectors by binarizing the query plan tree and encoding each query plan operator as a vector, optionally augmenting this representation with cache information.

\begin{figure}
  \centering
  \includegraphics[width=\linewidth]{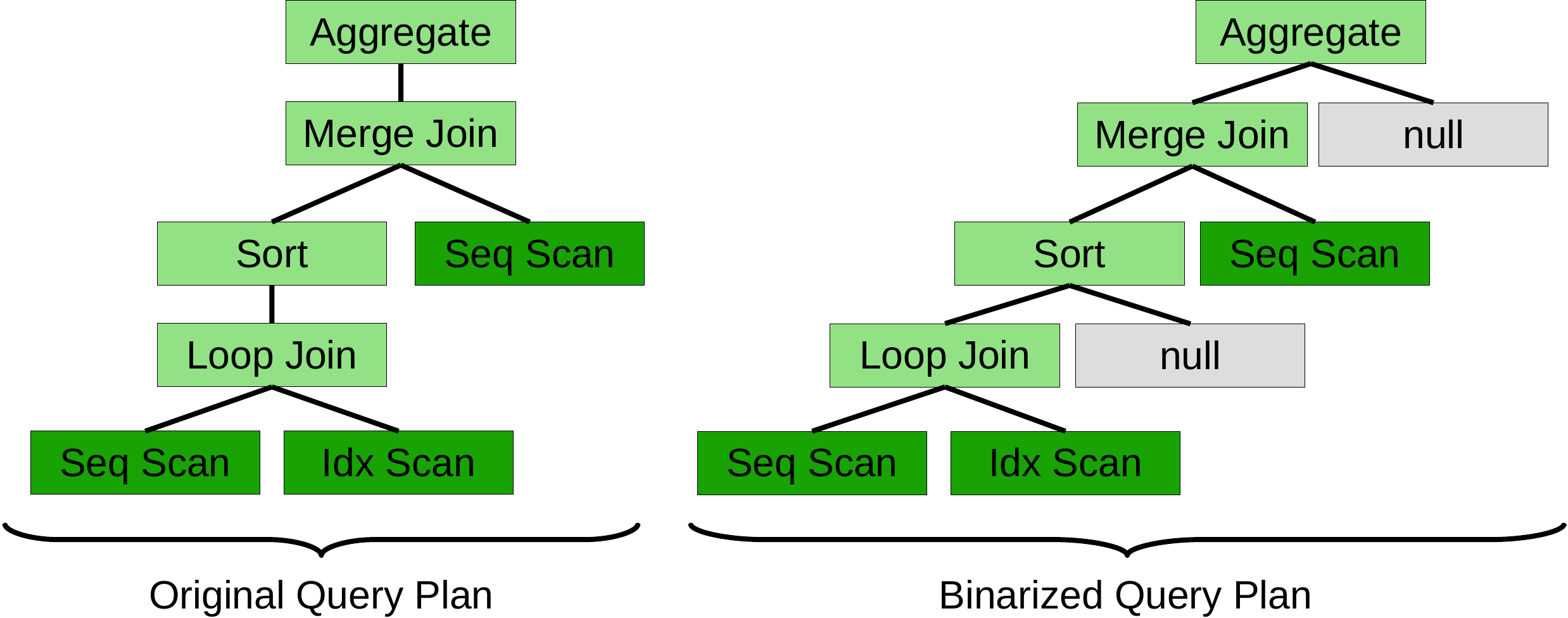}
  \caption{Binarizing a query plan tree}
  \label{fig:binary}
  \vspace{-4mm}
\end{figure}

\sparagraph{Binarization} Previous applications of reinforcement learning to query optimization~\cite{neo, rejoin, sanjay_wat} assumed that all query plan trees were strictly binary: every node in a query plan tree had either two children (an internal node) or zero children (a leaf). While this is true for a large class of simple join queries, most analytic queries involve non-binary operations like aggregation, sorting, and hashing. However, strictly binary query plan trees are convenient for a number of reasons, most notably that they greatly simplify tree convolution (explained in the next section). Thus, we propose a simple strategy to transform non-binary query plans into binary ones. Figure~\ref{fig:binary} shows an example of this process. The original query plan tree (left) is transformed into a binary query plan tree (right) by inserting ``null'' nodes (gray) as the right child of any node with a single parent. Nodes with more than two children (e.g., multi-unions) are uncommon, but can generally be binarized by splitting them up into a left-deep tree of binary operations (e.g., a union of 5 children is transformed into a left-deep tree with four binary union operators).

\begin{figure}
  \centering
  \includegraphics[width=\linewidth]{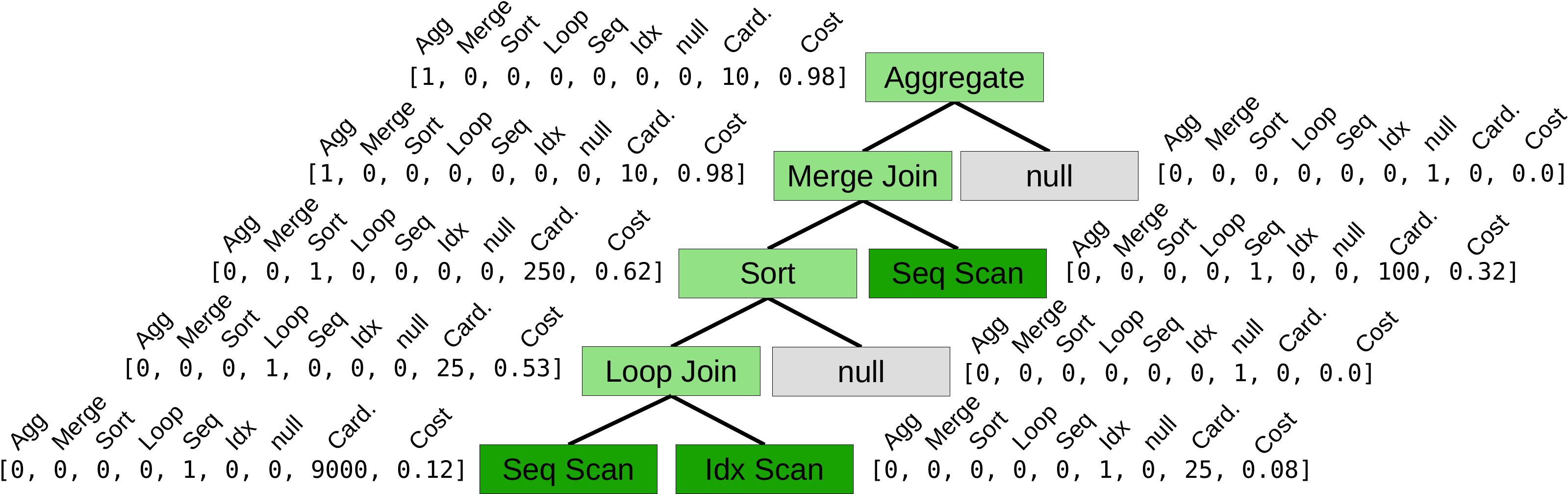}
  \caption{Vectorized query plan tree (vector tree)}
  \label{fig:vector}
\end{figure}

\sparagraph{Vectorization} Bao's vectorization strategy produces vectors with few components. Each node in a query plan tree is transformed into a vector containing three parts: (1) a one hot encoding of the operator type, (2) cardinality and cost model information, and optionally (3) cache information.

The one-hot encoding of each operator type is similar to vectorization strategies used by previous approaches~\cite{neo, sanjay_wat}. Each vector in Figure~\ref{fig:vector} begins with the one-hot encoding of the operator type (e.g., the second position is used to indicate if an operator is a merge join). This simple one-hot encoding captures information about structural properties of the query plan tree: for example, a merge join with a child that is not sort might indicate that more than one operator is taking advantage of a sorted order.

Each vector can also contain information about estimated cardinality and cost. Since almost all query optimizers  make use of such information, surfacing it to the vector representation of each query plan tree node is often trivial. For example, in Figure~\ref{fig:vector}, cardinality and cost model information is labeled ``Card'' and ``Cost'' respectively. This information helps encode if an operator is potentially problematic, such a loop joins over large relations or repetitive sorting, which might be indicative of a poor query plan. While we use only two values (one for a cardinality estimate, the other for a cost estimate), any number of values can be used. For example, multiple cardinality estimates from different estimators or predictions from learned cost models may be added.

Finally, optionally, each vector can be augmented with information from the current state of the disk cache. The current state of the cache can be retrieved from the database buffer pool when a new query arrives. In our experiments, we augment each scan node with the percentage of the targeted file that is cached, although many other schemes can be used. This gives Bao the opportunity to pick plans that are compatible with information in the cache.

While simple, Bao's vectorization scheme has a number of advantages. First, the representation is agnostic to the underlying schema: while prior work~\cite{sanjay_wat, rejoin, neo} represented tables and columns directly in their vectorization scheme, Bao omits them so that schema changes do not necessitate starting over from scratch. Second, Bao's vectorization scheme only represents the underlying data with cardinality estimates and cost models, as opposed to complex embedding models tied to the data~\cite{neo}. Since maintaining cardinality estimates when data changes is well-studied and already implemented in most DBMSes, changes to the underlying data are reflected cleanly in Bao's vectorized representation.

\subsubsection{Tree convolutional neural networks}
\label{sec:tcnn}
Tree convolution is a composable and differentiable neural network operator introduced in~\cite{tree_conv} for supervised program analysis and first applied to query plan trees in~\cite{neo}. Here, we give an intuitive overview of tree convolution, and refer readers to~\cite{neo} for technical details and analysis of tree convolution on query plan trees.


As noted in~\cite{neo}, human experts studying query plans learn to recognize good or bad plans by pattern matching: a pipeline of merge joins without any intermediate sorts may perform well, whereas a merge join on top of a hash join may induce a redundant sort or hash. Similarly, a hash join which builds a hash table over a very large relation may incur a spill. While none of this patterns are independently enough to decide if a query plan is good or bad, they do serve as useful indicators for further analysis; in other words, the presence or absence of such a pattern is a useful feature from a learning prospective.  Tree convolution is precisely suited to recognize such patterns, and learns to do so \emph{automatically, from the data itself}.

Tree convolution consists of sliding several tree-shaped ``filters'' over a larger query plan tree (similar to image convolution, where filters in a filterbank are convolved with an image) to produce a transformed query plan tree of the same size. These filters may look for patterns like pairs of hash joins, or an index scan over a very small relation. Tree convolution operators are stacked, resulting in several layers of tree convolution. Later layers can learn to recognize more complex patterns, like a long chain of merge joins or a bushy tree of hash operators. Because of tree convolution's natural ability to represent and learn these patterns, we say that tree convolution represents a helpful \emph{inductive bias}~\cite{inductive_bias_rl, inductive_bias_ml} for query optimization: that is, the structure of the network, not just its parameters, are tuned to the underlying problem.

\begin{figure}
  \centering
  \includegraphics[width=0.48\textwidth]{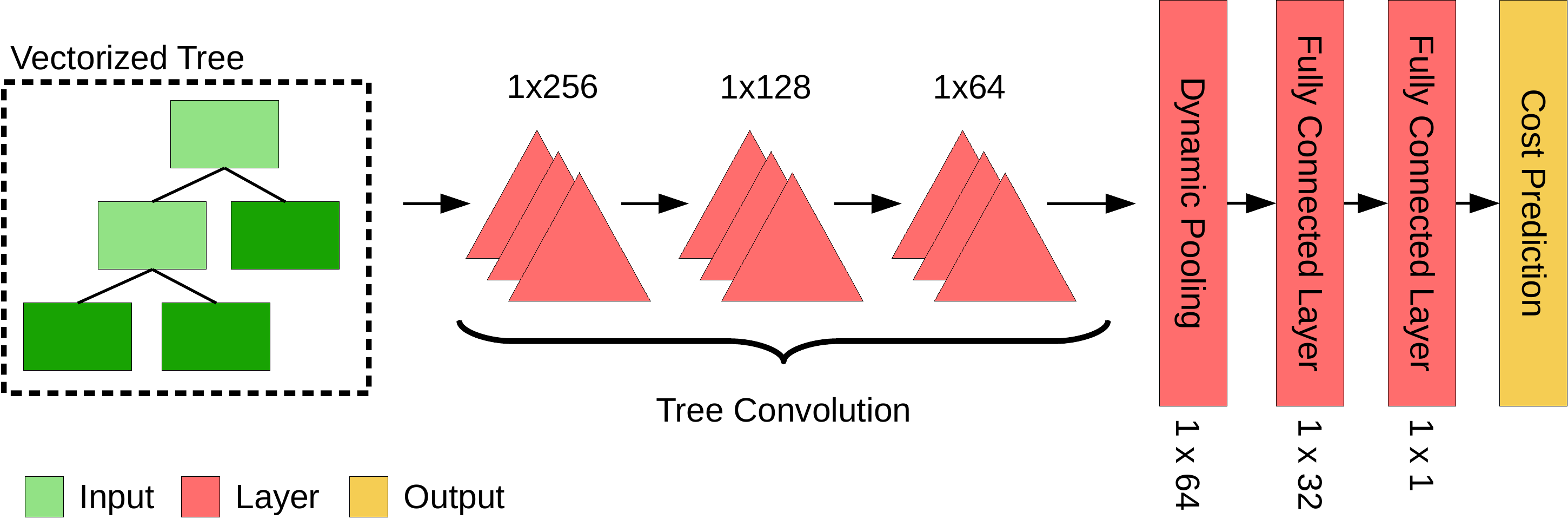}
  \vspace{1mm}
  \caption{Bao prediction model architecture}
  \label{fig:network}
\end{figure}

The architecture of Bao's prediction model (similar to Neo's value prediction model~\cite{neo}) is shown in Figure~\ref{fig:network}. The vectorized query plan tree is passed through three layers of stacked tree convolution. After the last layer of tree convolution, dynamic pooling~\cite{tree_conv} is used to flatten the tree structure into a single vector. Then, two fully connected layers are used to map the pooled vector to a performance prediction. We use ReLU~\cite{relu} activation functions and layer normalization~\cite{layer_norm}, which are not shown in the figure.

\sparagraph{Integrating with Thompson sampling} Thompson sampling requires the ability to \emph{sample} model parameters $\theta$ from $P(\theta \mid E)$, whereas most machine learning techniques are designed to find the most likely model given the training data, $\mathbb{E}[P(\theta \mid E)]$. For neural networks, there are several techniques available to sample from $P(\theta \mid E)$, ranging from complex Bayesian neural networks to simple approaches~\cite{deep_bayes_bandits}. By far the simplest technique, which has been shown to work well in practice~\cite{thompson_bootstrap}, is to train the neural network as usual, but on a ``bootstrap''~\cite{bootstrapping} of the training data: the network is trained using $|E|$ random samples drawn with replacement from $E$, inducing the desired sampling properties~\cite{thompson_bootstrap}. We selected this bootstrapping technique for its simplicity.

\subsection{Training loop}
\label{sec:train_loop}

In this section, we explain Bao's training loop, which closely follows a classical Thompson sampling regime: when a query is received, Bao  builds a query plan tree for each hint set and then uses the current TCNN predictive model to select a query plan tree to execute. After execution, that query plan tree and the observed performance are added to Bao's experience. Periodically, Bao retrains its TCNN predictive model by sampling model parameters (i.e., neural network weights) to balance exploration and exploitation. While Bao closely follows a Thompson sampling regime for solving a contextual multi-armed bandit, practical concerns require a few deviations.

In classical Thompson sampling~\cite{thompson}, the model parameters $\theta$ are resampled after every selection (query). In the context of query optimization, this is not practical for two reasons. First, sampling $\theta$ requires training a neural network, which is a time consuming process. Second, if the size of the experience $|E|$ grows unbounded as queries are processed, the time to train the neural network will also grow unbounded, as the time required to perform a training epoch is linear in the number of training examples.

We use two techniques from prior work on using deep learning for contextual multi-armed bandit problems~\cite{deep_cmab} to solve these issues. First, instead of resampling the model parameters (i.e., retraining the neural network) after every query, we only resample the parameters every $n$th query. This obviously decreases the training overhead by a factor of $n$ by using the same model parameters for more than one query. Second, instead of allowing $|E|$ to grow unbounded, we only store the $k$ most recent experiences in $E$. By tuning $n$ and $k$, the user can control the tradeoff between model quality and training overhead to their needs. We evaluate this tradeoff in Section~\ref{sec:warm}.

We also introduce a new optimization, specifically useful for query optimization. On modern cloud platforms such as~\cite{url-google}, GPUs can be attached and detached from a VM with per-second billing. Since training a neural network primarily uses the GPU, whereas query processing primarily uses the CPU, disk, and RAM, model training and query execution can be overlapped. When new model parameters need to be sampled, a GPU can be temporarily provisioned and attached. Model training can then be offloaded to the GPU, leaving other resources available for query processing. Once model training is complete, the new model parameters can be swapped in for use when the next query arrives, and the GPU can be detached. Of course, users may also choose to use a machine with a dedicated GPU, or to offload model training to a different machine entirely.


\section{Related work}
\label{sec:rw}

Recently, there has been a groundswell of research on integrating machine learning into query optimization. One of the most obvious places in query optimization to apply machine learning is cardinality estimation. One of the earliest approaches was Leo~\cite{leo}, which used successive runs of the similar queries to adjust histogram estimators. More recent approaches~\cite{deep_card_est, deep_card_est2, local_card_est, learn_cost} have used deep learning to learn cardinality estimations or query costs in a supervised fashion, although these works require extensive training data collection and do not adapt to changes in data or schema. QuickSel~\cite{quicksel} demonstrated that linear mixture models could learn reasonable estimates for single tables. Naru~\cite{naru} uses an unsupervised learning approach which does not require training and uses Monte Carlo integration to produce estimates, again for materialized tables.  In~\cite{containment_rates}, authors present a scheme called CRN for estimating cardinalities via query containment rates. While all of these  works demonstrate improved cardinality estimation accuracy, they do not provide  evidence that these improvements lead to better query performance. In an experimental study, Ortiz et al.~\cite{learned_card_eval} show that certain learned cardinality estimation techniques may improve mean performance on certain datasets, but tail latency is not evaluated. In~\cite{plan_loss}, Negi et al. show how prioritizing training on cardinality estimations that have a large impact on query performance can improve estimation models.

Another line of research has examined using reinforcement learning to construct query optimizers. Both~\cite{rejoin, sanjay_wat} showed that, with sufficient training, such approaches could find plans with lower costs according to the PostgreSQL optimizer and cardinality estimator. \cite{qo_state_rep} showed that the internal state learned by reinforcement learning algorithms are strongly related to cardinality. Neo~\cite{neo} showed that deep reinforcement learning could be applied directly to query latency, and could learn optimization strategies that were competitive with commercial systems after 24 hours of training. However, none of these techniques are capable of handling changes in schema, data, or query workload. Furthermore, while all of these techniques show improvements to \emph{mean} query performance after a long training period, none demonstrate improvement in \emph{tail} performance.

Works applying reinforcement learning to adaptive query processing~\cite{skinnerdb, cuttlefish, adaptive_qp_rl} have also shown interesting results, but are not applicable to non-adaptive systems.

Thompson sampling has a long history in statistics and decision making problems and recently it has been used extensively in the reinforcement learning community as a simple yet efficient way to update beliefs given experience~\cite{thompson,thompson_bound_time,thompson_intro}. We use an alternative setup which lets us get the benefits of Thompson sampling without explicitly defining how to update the posterior belief, as described in~\cite{thompson_bootstrap}. Thompson sampling has also been applied to cloud workload management~\cite{wisedb-cidr} and SLA conformance~\cite{slaorchestrator}.

Reinforcement learning techniques in general have also seen recent adoption~\cite{lift}. In~\cite{sagedb}, the authors present a vision of an entire database system built from reinforcement learning components. More concretely, reinforcement learning has been applied to managing elastic clusters~\cite{mdp_elastic, perfenforce}, scheduling~\cite{deep_schedule}, and physical design~\cite{selfdrivingcidr}. 

Our work is part of a recent trend in seeking to use machine learning to build easy to use, adaptive, and inventive systems, a trend more broadly known as machine programming~\cite{pillars}. A few selected works outside the context of data management systems include reinforcement learning for job scheduling~\cite{decima}, automatic performance analysis~\cite{autoperf}, loop vectorization~\cite{neurovec} and  garbage collection~\cite{gc_rf}.



\begin{figure}
\centering
\begin{subfigure}{0.48\textwidth}
    \includegraphics[width=\textwidth]{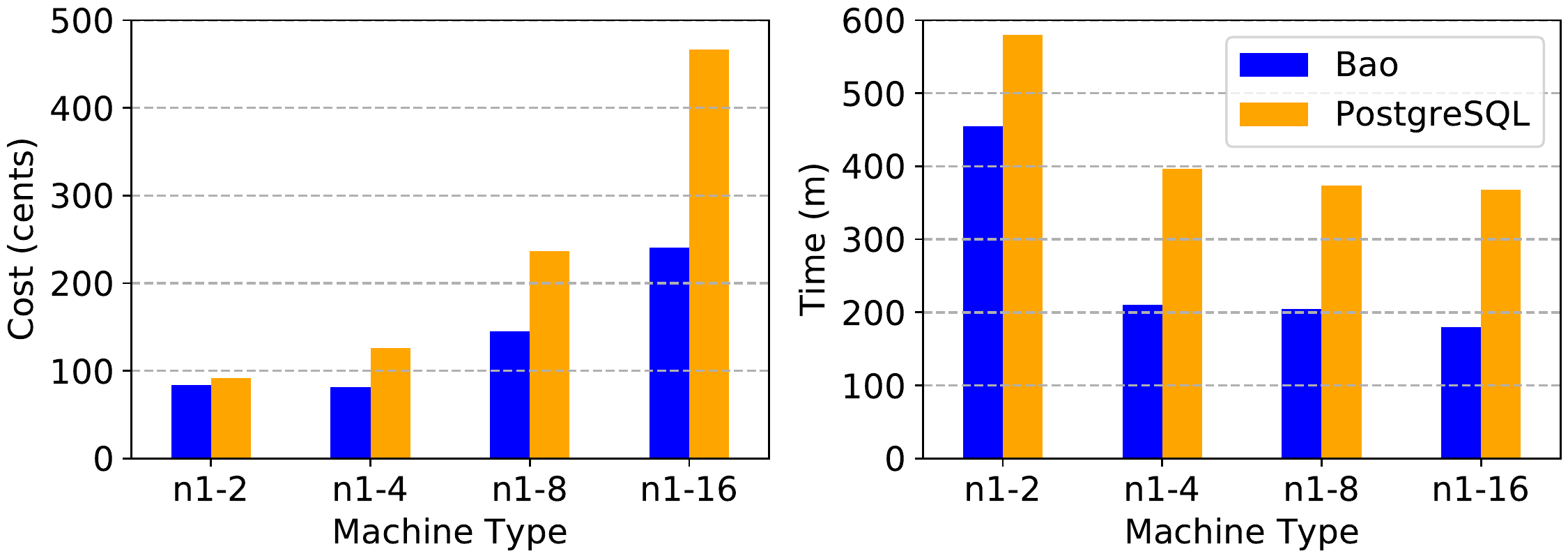}
    \caption{Across four different VM types, Bao on the PostgreSQL engine vs. PostgreSQL optimizer on the PostgreSQL engine.}
    \label{fig:pg_cvp}
\end{subfigure} 
\begin{subfigure}{0.48\textwidth}
    \includegraphics[width=\textwidth]{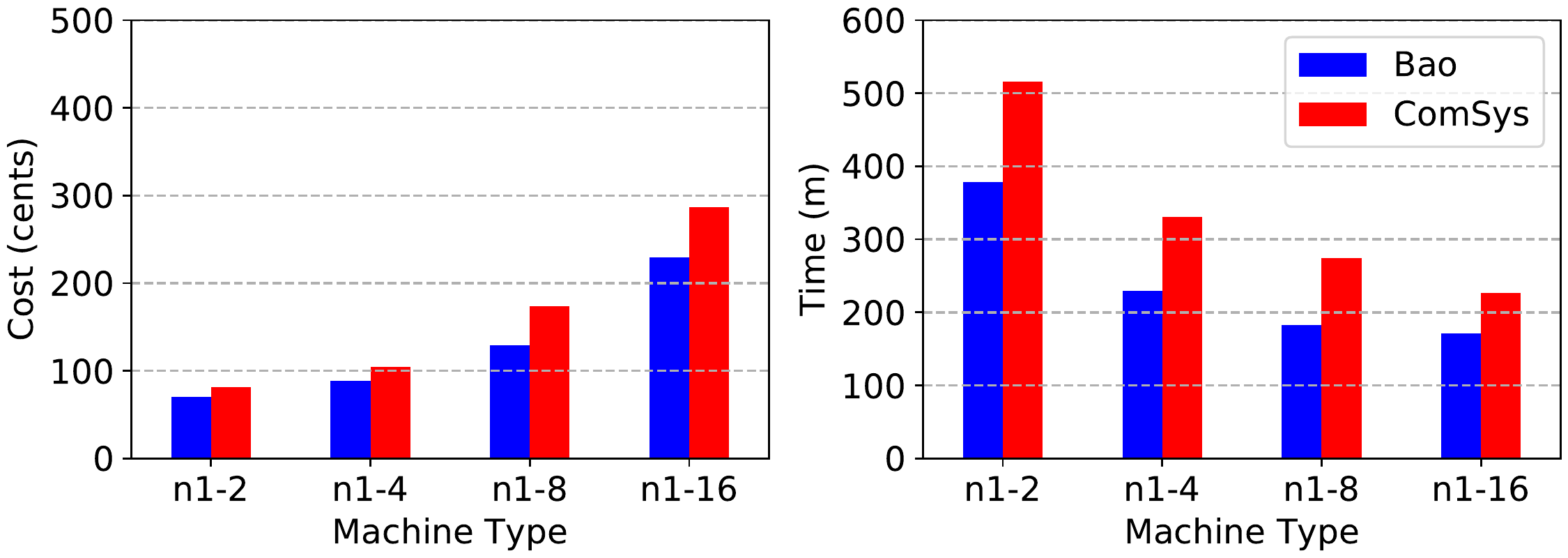}
    \caption{Across four different VM types, Bao on the ComSys engine vs. ComSys optimizer on the ComSys engine.}
    \label{fig:comsys_cvp}
  \end{subfigure} 
\caption{Cost (left) and workload latency (right) for Bao and two traditional query optimizers across four different Google Cloud Platform VM sizes for the IMDb workload.}
\label{fig:cvp}
\end{figure}

\begin{figure}
\centering
\begin{subfigure}{0.48\textwidth}
    \includegraphics[width=\textwidth]{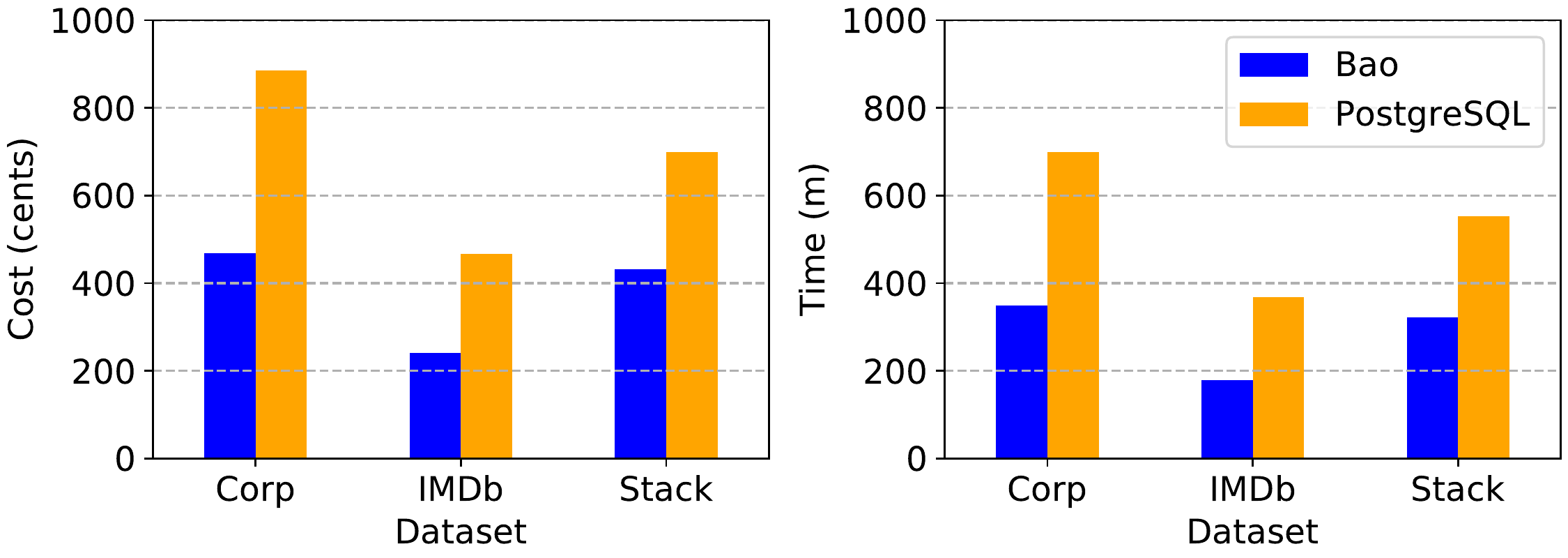}
    \caption{Across our three evaluation datasets, Bao on the PostgreSQL engine vs. PostgreSQL optimizer on the PostgreSQL engine.}
    \label{fig:pg_cvp_ds}
\end{subfigure} 
\begin{subfigure}{0.48\textwidth}
    \includegraphics[width=\textwidth]{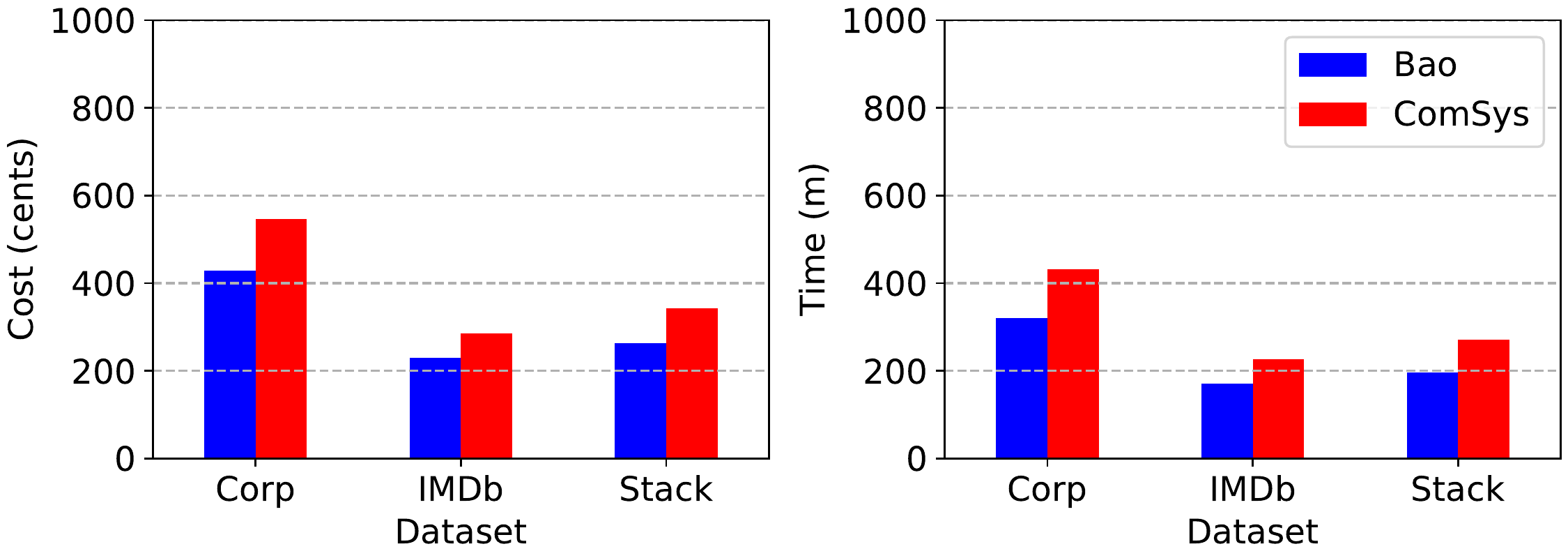}
    \caption{Across our three evaluation datasets, Bao on the ComSys engine vs. ComSys optimizer on the ComSys engine.}
    \label{fig:comsys_cvp_ds}
  \end{subfigure} 
\caption{Cost (left) and workload latency (right) for Bao and two traditional query optimizers across three different workloads on a N1-16 Google Cloud VM.}
\label{fig:cvp_ds}
\end{figure}

\section{Experiments}
\label{sec:expr}
\hl{The key question we pose in our evaluation is whether or not Bao could have a positive, practical impact on real-world database workloads that include changes in workload, data, and/or schema. To answer this, we focus on quantifying not only query performance, but also on the dollar-cost of executing a workload (including the training overhead introduced by Bao) on cloud infrastructure.}

Our experimental study is divided into three parts. In Section~\ref{sec:expr_setup}, we explain our experimental setup. Section~\ref{sec:warm} is designed to evaluate Bao's real-world applicability, and compares Bao's performance against both PostgreSQL and a commercial database system~\cite{dewitt_clause} on real-world workloads executed on Google Cloud Platform with caching enabled. Specifically, we examine:

\begin{itemize}
\item{workload performance, tail latency, and total costs for executing workloads with Bao, the PostgreSQL optimizer, and a commercial system with dynamic workloads, data, and schema,}
\item{Bao's resiliency to hints that induce consistent or sudden poor performance,}
\item{quantitative and qualitative differences between Bao and  previous learned query optimization approaches.}
\end{itemize}

Section~\ref{sec:cold} is designed to evaluate Bao's optimality in terms of \emph{regret}, the difference between Bao's decisions and the optimal choice. Experiments in these sections are necessarily conducted on a private, isolated server, and each query is executed with a cold cache. \hl{This section evaluates:

\begin{itemize}
\item{Bao's ability to adapt to different optimization goals (e.g., disk IOs),}
\item{the median and tail latency achieved by Bao relative to an optimal choice and an open source query optimizer,}
\item{Bao's query-by-query performance on the join order benchmark~\cite{howgood}.}
\end{itemize}}

\subsection{Setup}
\label{sec:expr_setup}

\begin{table}
\begin{minipage}{0.48\textwidth}
\renewcommand\footnoterule{}
\centering
\begin{tabularx}{\textwidth}{lXXXXX}
\toprule
  & {\bf Size} & {\bf Queries} & {\bf WL} & {\bf Data} & {\bf Schema} \\
\midrule
{\bf IMDb}          & 7.2  GB & 5000      & Dynamic & Static & Static\\
{\bf Stack}         & 100  GB & 5000      & Dynamic & Dynamic & Static\\
{\bf Corp}          & 1 TB    & 2000      & Dynamic & Static\footnote{The schema change did not introduce new data, but did normalize a large fact table. Thus, we consider the data static.} & Dynamic  \\
\bottomrule
\end{tabularx}
\end{minipage}
\caption{Evaluation dataset sizes, query counts, and whether or not the workload (WL), data, and schema are static or dynamic.}
\label{tbl:datasets}
\end{table}

We evaluated Bao using the datasets listed in Table~\ref{tbl:datasets}.
\begin{itemize}
\item{The IMDb dataset is an augmentation of the Join Order Benchmark~\cite{howgood}: we added thousands of queries to the original 113 queries,\footnote{\url{https://rm.cab/imdb}} and we vary the query workload over time by introducing new templates periodically. The data and schema remain static.}
\item{The Stack dataset is a new dataset introduced by this work and available publicly.\footnote{\url{https://rm.cab/stack}} The Stack dataset contains over 18 million questions and answers from 170 different StackExchange websites (such as StackOverflow.com) between July 2008 and September 2019. We emulate data drift by initially loading all data up to September 2018, and then incrementally inserting the data from September 2018 to September 2019. We have produced 5000 queries from 25 different templates, and we vary the query workload by introducing new templates periodically. The schema remains static.}
\item{The Corp dataset is a dashboard analytics workload executed over one month donated by an anonymous corporation. The Corp dataset contains 2000 unique queries issued by analysts. Half way through the month, the corporation normalized a large fact table, resulting in a significant schema change. We emulate this schema change by introducing the normalization after the execution of the 1000th query (queries after the 1000th expect the new normalized schema). The data remains static.}
\end{itemize}

For the Stack and IMDb workloads, we vary the workload over time by introducing new query templates periodically. We choose the query sequence by randomly assigning each query template to two of eight groups (every query template is in exactly two groups). We then build 8 corresponding groups of query instances, placing one half of all instances of a query template into the corresponding query template group. We then randomly order the query instances within each group, and concatenate the groups together to determine the order of the queries in the workload. This ensures that a wide variety of template combinations and shifts are present. Note that for the Corp workload, the queries are replayed in the same order as analysts issues them.

Unless otherwise noted, we use a ``time series split'' strategy for training and testing Bao. Bao is always evaluated on the next, never-before-seen query $q_{t+1}$. When Bao makes a decision for query $q_{t+1}$, Bao is only trained on data from earlier queries. Once Bao makes a decision for query $q_{t+1}$, the observed reward for that decision -- and only that decision -- is added to Bao's experience set. This strategy differs from previous evaluations in~\cite{neo, rejoin, sanjay_wat, qo_state_rep} because Bao is never allowed to learn from two different decisions about the same query.\footnote{\hl{In OLAP workloads where nearly-identical queries are frequently repeated (e.g., dashboards), this may be an overcautious procedure.}} Whereas in the prior works mentioned, a reinforcement learning algorithm could investigate many possible decisions about the same query, our technique is a more realistic: once a query is executed using a particular plan, Bao does not get any information about alternative plans for the same query.

Bao's prediction model uses three layers of tree convolution, with output dimensions (256, 128, 64), followed by a dynamic pooling~\cite{tree_conv} layer and two linear layers with output dimensions (32, 1). We use ReLU activation functions~\cite{relu} and layer normalization~\cite{layer_norm} between each layer. Training is performed with Adam~\cite{adam} using a batch size of 16, and is ran until either 100 epochs elapsed or convergence is reached (as measured by a decrease in training loss of less than 1\% over 10 epochs).

Experiments in Section~\ref{sec:warm} are performed on Google Cloud Platform, using a N1-4 VM type unless otherwise noted. NVIDIA Tesla T4 GPUs are attached to VMs when needed. Cost and time measurements include query optimization, model training (including GPU time), and query execution. Costs are reported as billed by Google, and include startup times and minimum usage thresholds. Experiments in Section~\ref{sec:cold} are performed on a virtual machine with 4 CPU cores and 15 GB of RAM (to match the N1-4 VMs) on private server with two Intel(R) Xeon(R) Gold 6230 CPUs running at 2.1 Ghz, an NVIDIA Tesla T4 GPU, and 256GB of system (bare metal) RAM.

We compare Bao against the open source PostgreSQL database and a commercial database system (ComSys) we are not permitted to name~\cite{dewitt_clause}. Both systems are configured and tuned according to their respective documentation and best practices guide. A consultant from the company that produces the commercial system under test verified our configuration with a small performance test. Bao's chosen execution plan is always executed on the system being compared against: for example, when comparing against the PostgreSQL optimizer, Bao's execution plans are always executed on the PostgreSQL engine.

Unless otherwise noted, we use a family of 48 hint sets, which each use some subset of the join operators \{hash join, merge join, loop join\} and some subset of the scan operators \{sequential, index, index only\}. For a detailed description, see the online appendix~\cite{bao_arxiv}. We found that setting the lookback window size to $k = 2000$ and retraining every $n = 100$ queries provided a good tradeoff between GPU time and query performance.

\subsection{Real-world performance}
\label{sec:warm}

In this section, we evaluate Bao's performance in a realistic warm-cache scenario. In these experiments, we augment each leaf node vector with caching information as described in Section~\ref{sec:vectorize}. For caching information, we queried the commercial system's internal system tables to determine what percentage of each file was cached. We used a similar strategy for PostgreSQL, using the built-in \texttt{pg\_buffercache} extension in the \texttt{contrib} folder.

\begin{figure*}[t]
  \begin{tabular}{m{1mm} m{0.22\textwidth} m{0.22\textwidth} m{0.22\textwidth} m{0.22\textwidth}}
    & \centering N1-2 & \centering N1-4 & \centering N1-8 & \begin{center}N1-16\end{center} \\
    \rotatebox{90}{PostgreSQL}
    & \includegraphics[width=0.22\textwidth]{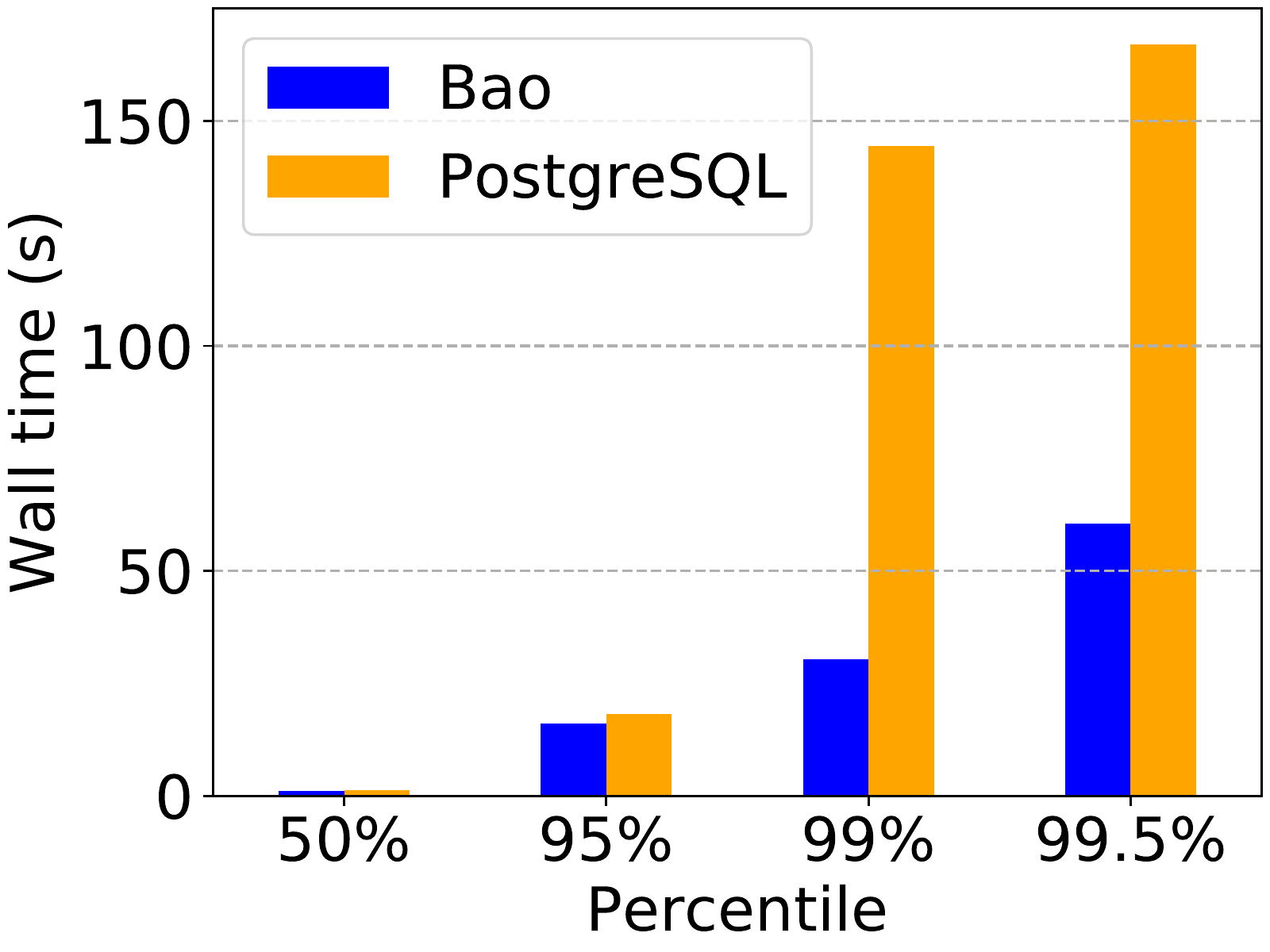}
    & \includegraphics[width=0.22\textwidth]{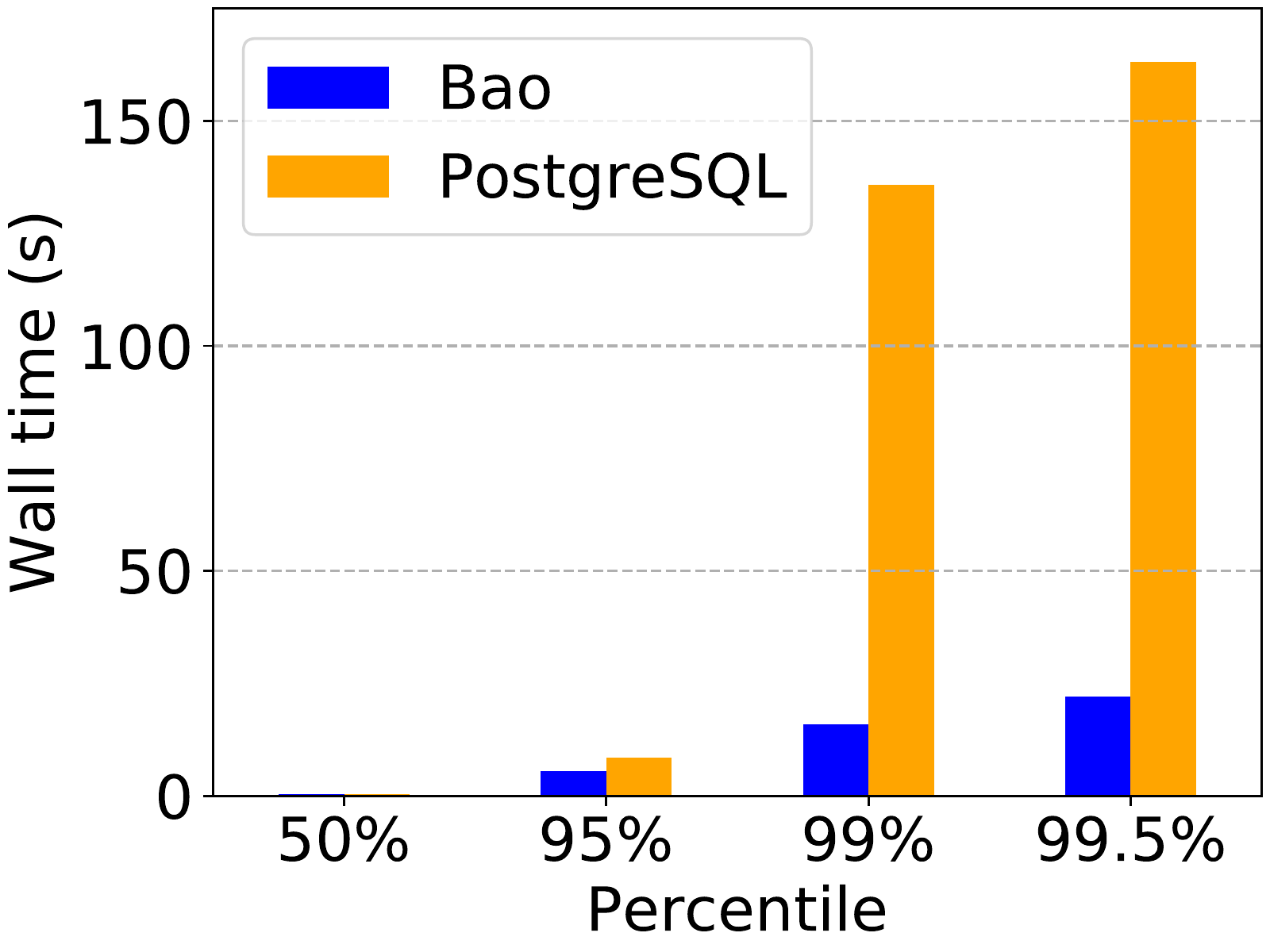}
    & \includegraphics[width=0.22\textwidth]{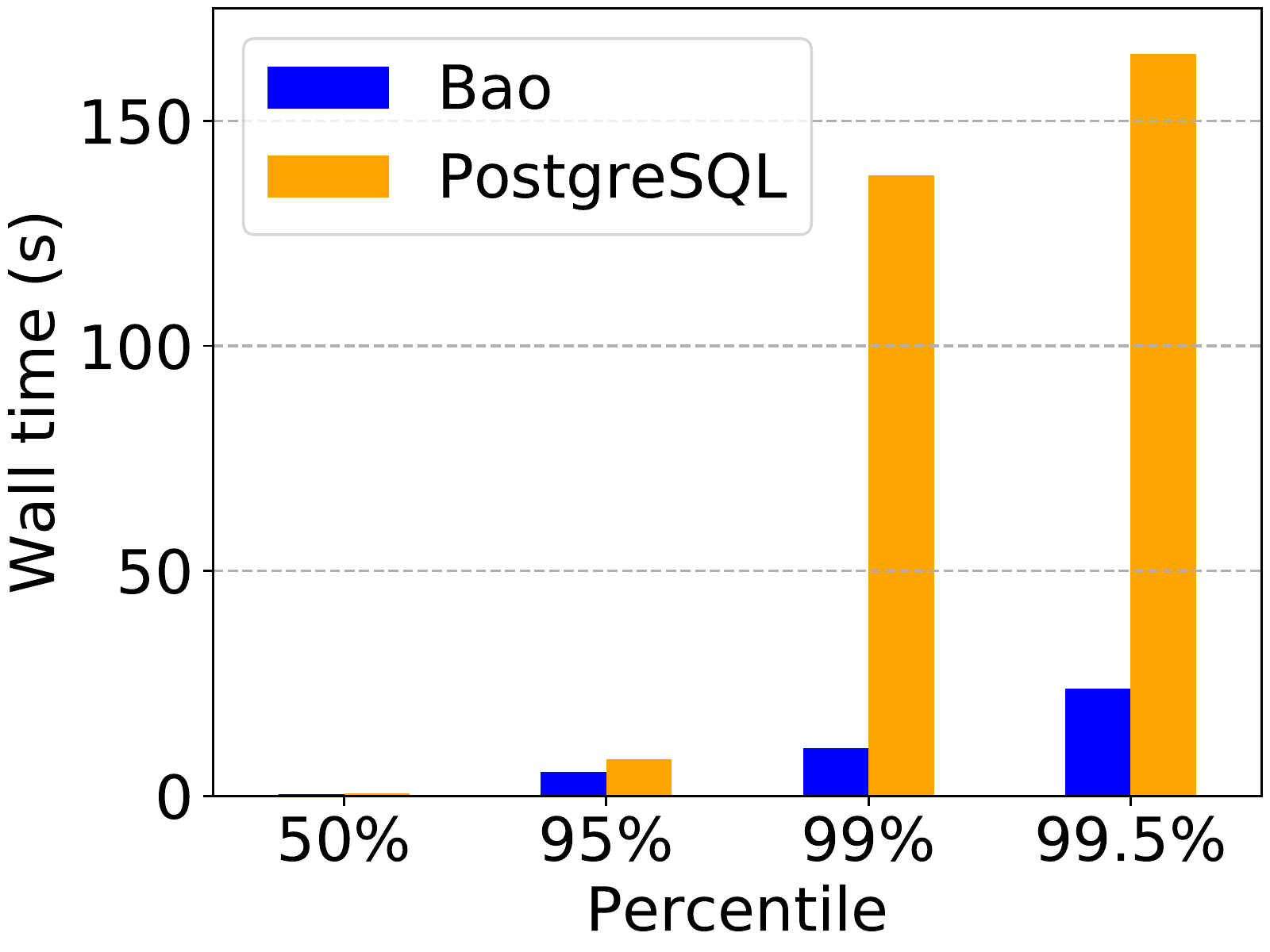}
    & \includegraphics[width=0.22\textwidth]{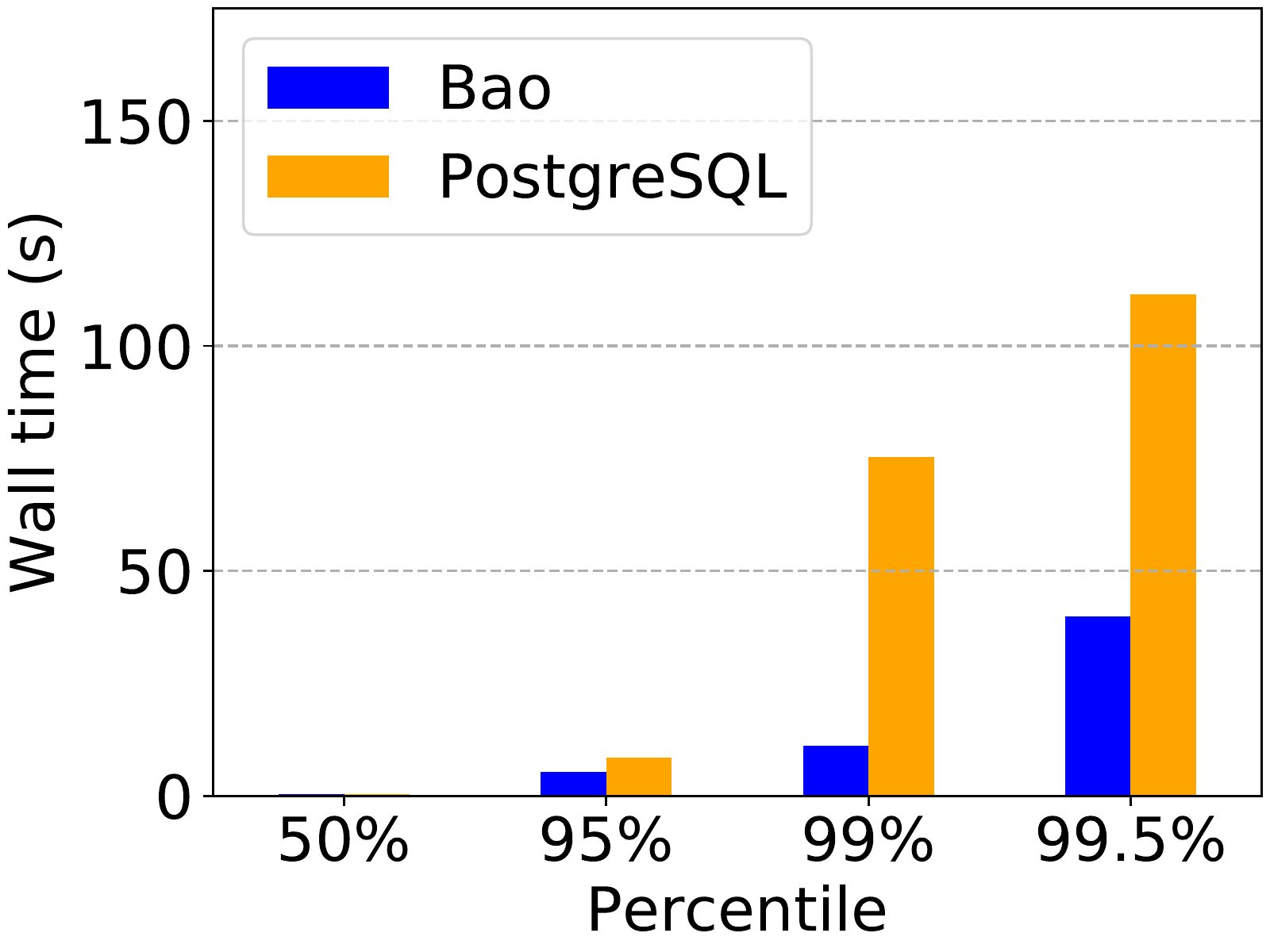} \\
    \rotatebox{90}{ComSys}
    & \includegraphics[width=0.22\textwidth]{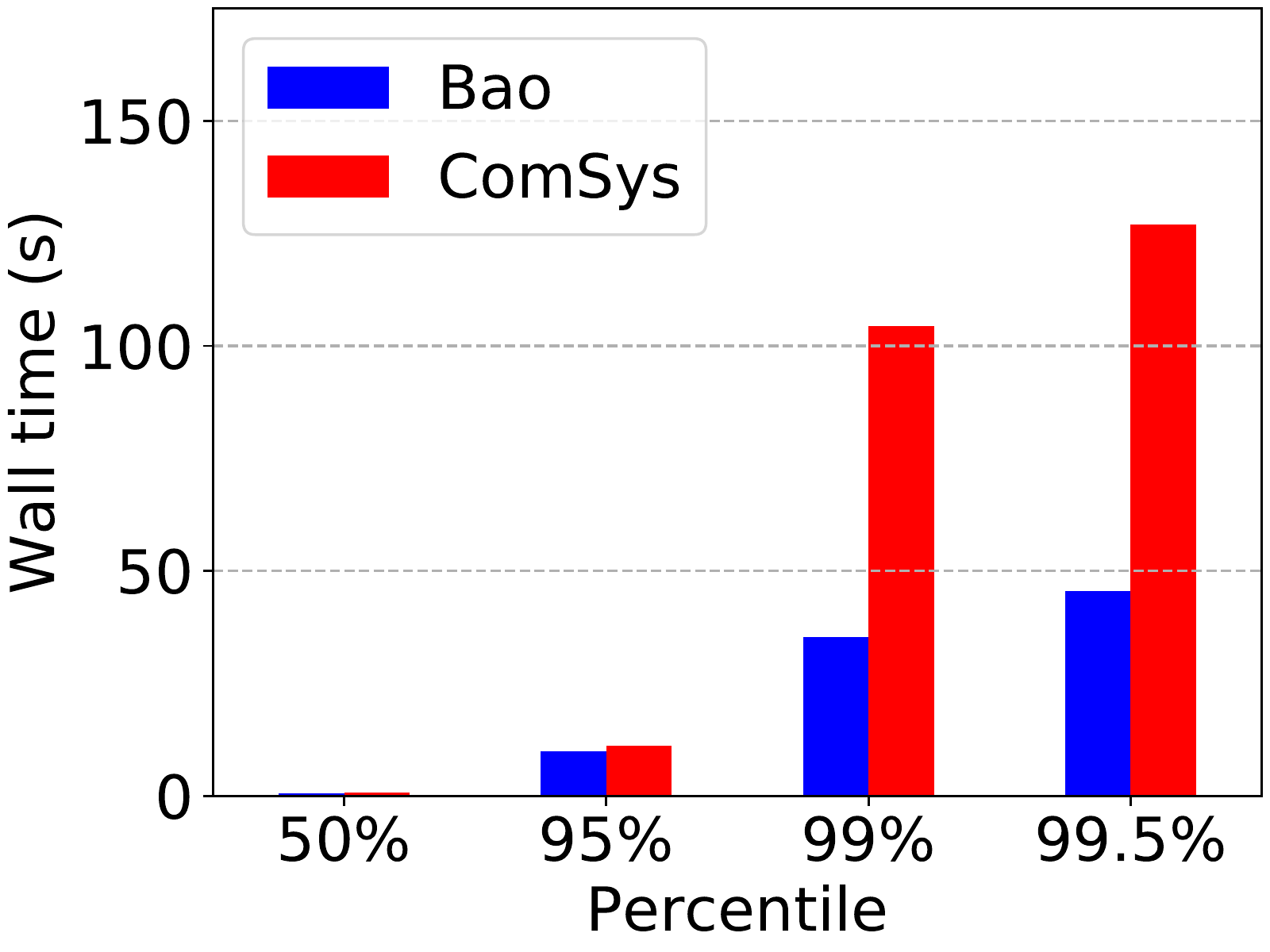}
    & \includegraphics[width=0.22\textwidth]{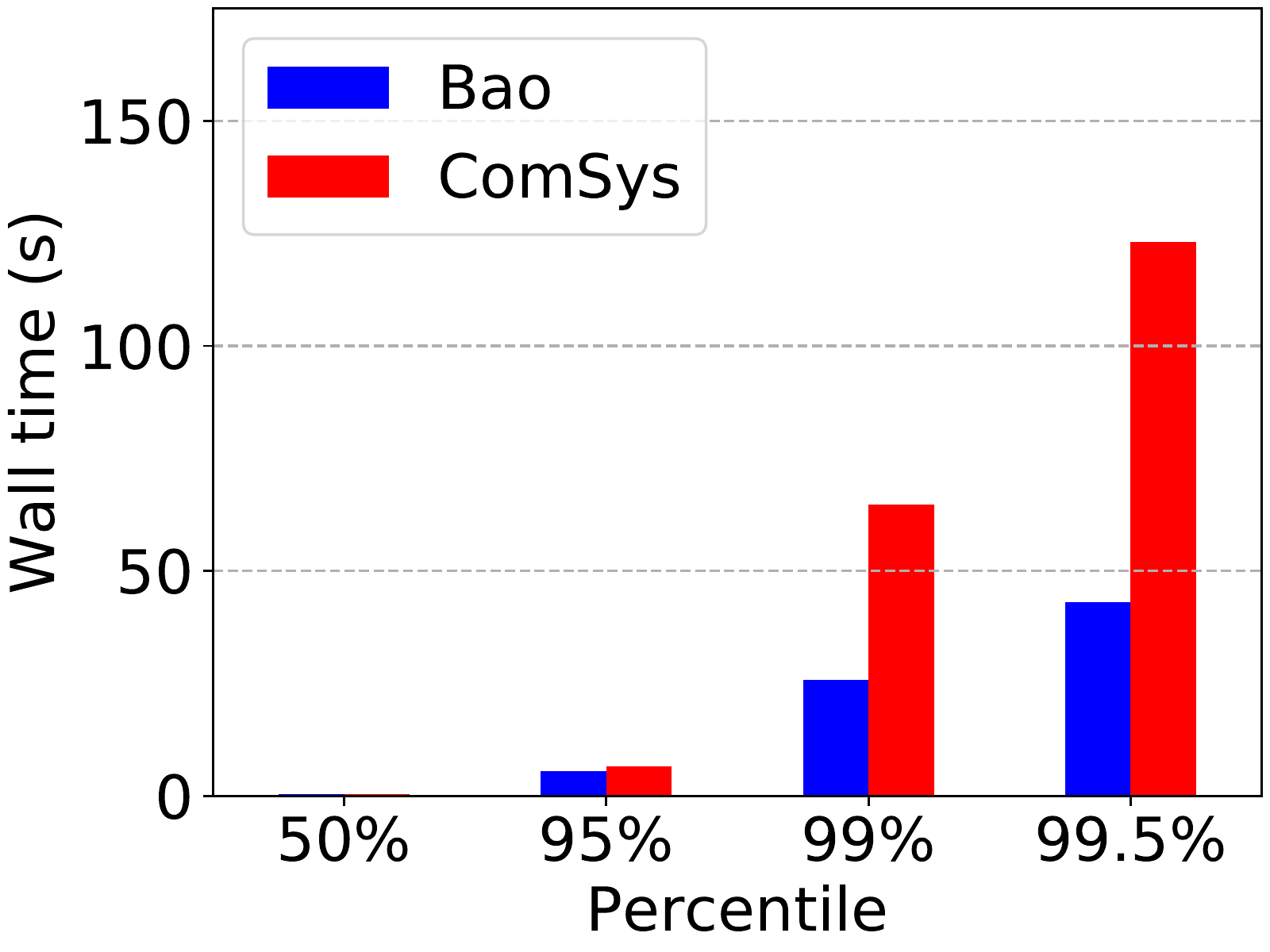}
    & \includegraphics[width=0.22\textwidth]{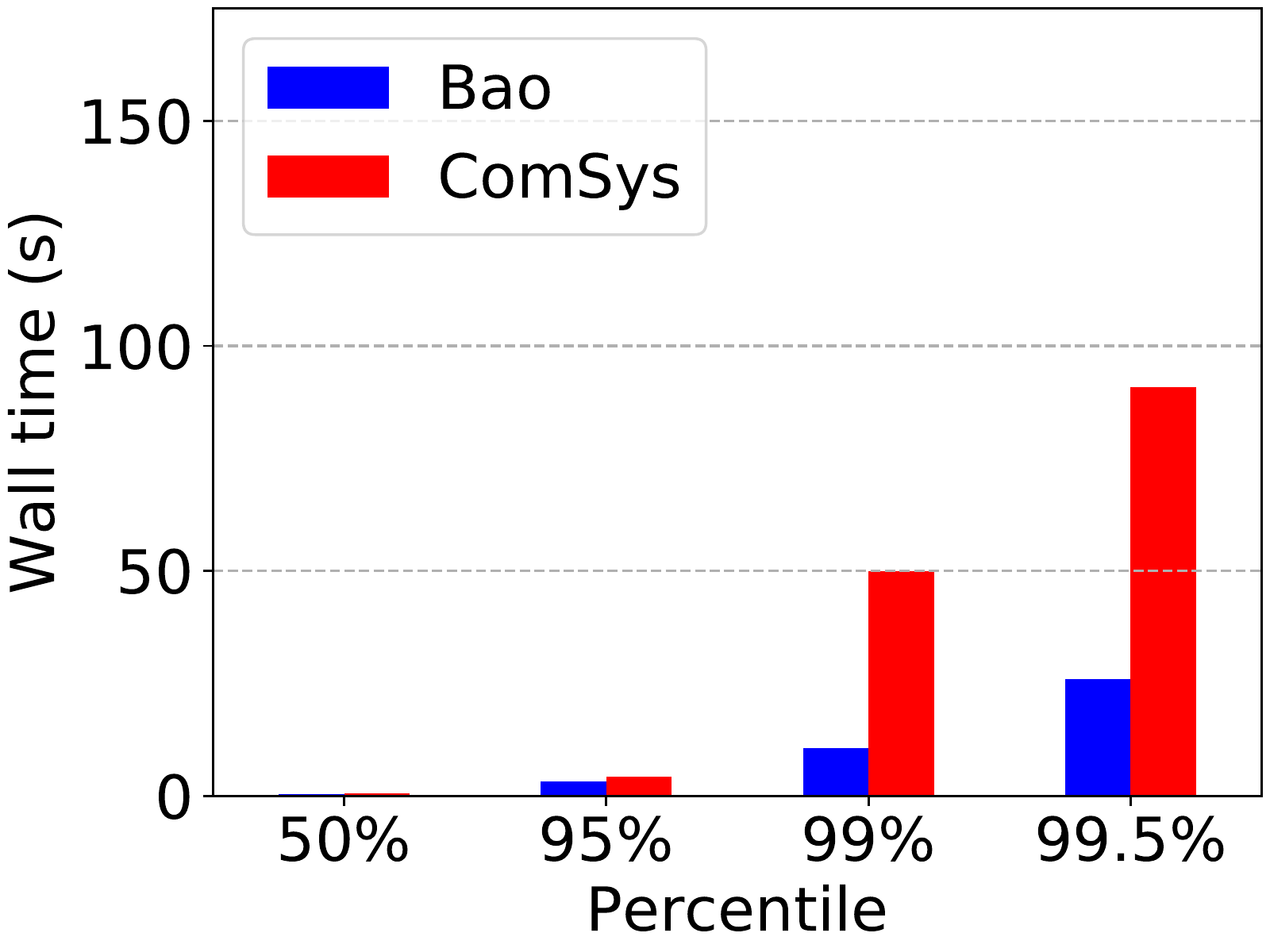}
    & \includegraphics[width=0.22\textwidth]{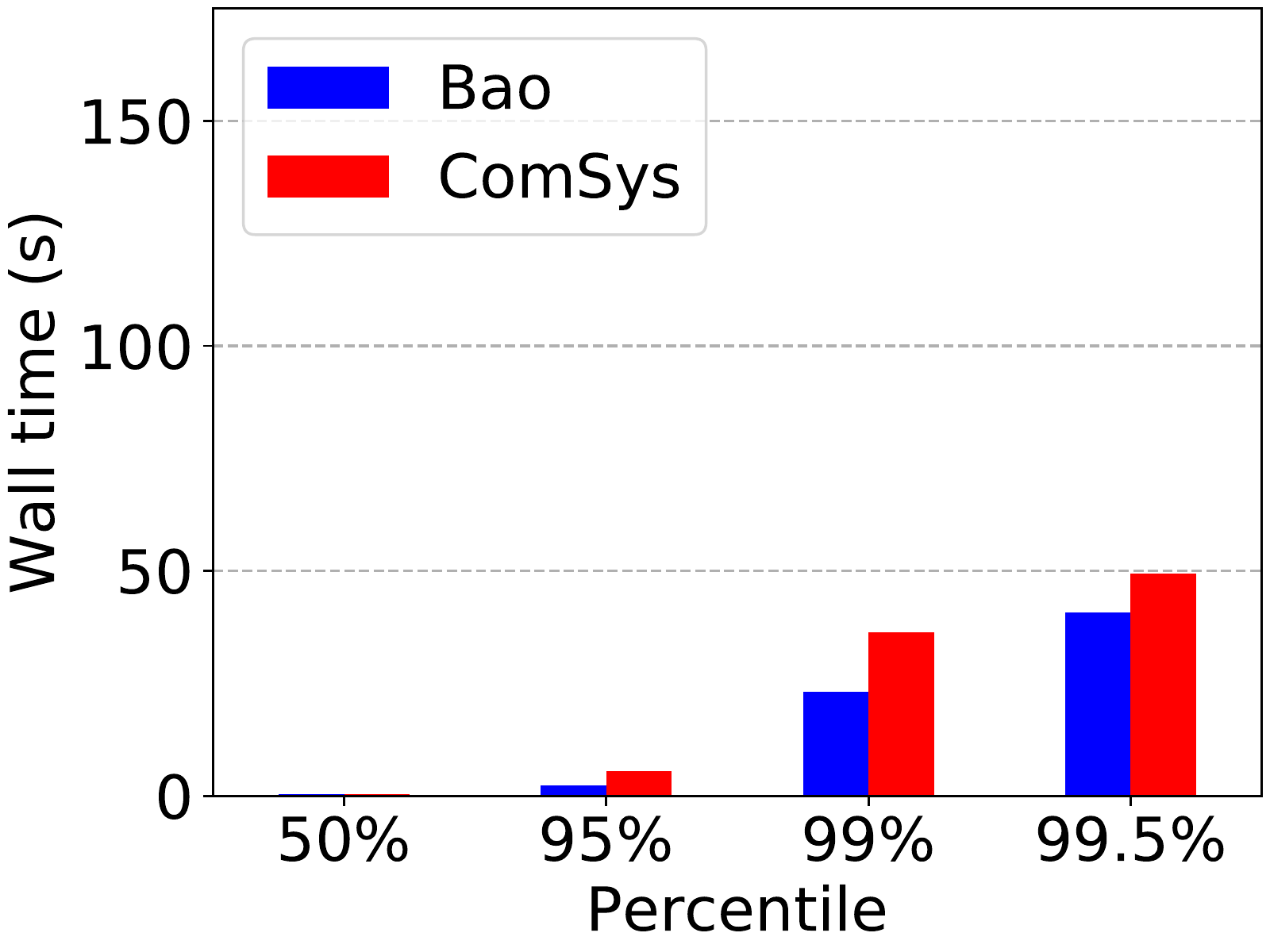} \\
  \end{tabular}
  \caption{Percentile latency for queries, IMDb workload. Each column represents a VM type, from smallest to largest. The top row compares Bao against the PostgreSQL optimizer on the PostgreSQL engine. The bottom row compares Bao against a commercial database system on the commercial system's engine. \hl{Measured across the entire (dynamic) IMDb workload.}}
  \label{fig:curves}
\end{figure*}

\sparagraph{Cost and performance in the cloud} To evaluate  Bao's potential impact on both query performance and cost, we begin by evaluating Bao on the Google Cloud Platform~\cite{url-google}. Figure~\ref{fig:pg_cvp} shows the cost (left) and time required (right) to execute the IMDb workload on various VM sizes when using Bao and when using the PostgreSQL optimizer on the PostgreSQL engine. Generally, \emph{Bao achieves both a lower cost and a lower workload latency.} For example, on a N1-16 VM, Bao reduces costs by over 50\% (from \$4.60 to \$2.20) while also reducing the total workload time from over six hours to just over three hours. The Bao costs \emph{do} include the additional fee for renting a GPU: the increased query performance more than makes up for the cost incurred from attaching a GPU for training. The difference in both cost and performance is most significant with larger VM types (e.g., N1-16 vs. N1-8), suggesting the Bao is better capable of tuning itself towards the changing hardware than PostgreSQL. We note that we \emph{did} re-tune PostgreSQL for each hardware platform.

Figure~\ref{fig:comsys_cvp} shows the same comparison against the commercial database system. Again, Bao is capable of achieving lower cost and lower workload latency on all four tested machine types. However, the difference is less significant, and the overall costs are much lower, suggesting the commercial system is a stronger baseline than PostgreSQL. For example, while Bao achieved almost a 50\% cost and latency reduction on the N1-16 machine compared to the PostgreSQL optimizer, Bao achieves only a 20\% reduction compared to the commercial system. We also note that the improvements from Bao are no longer more significant with larger VM types, indicating that the commercial system is more capable of adjusting to different hardware. Note that these costs do \emph{not} include the licensing fees for the commercial system, which were waived for the purposes of this research.

\sparagraph{\hl{Changing schema, workload, and data}} In Figure~\ref{fig:cvp_ds}, we fix the VM type to N1-16 and evaluate Bao on different workloads. Bao shows significant improvements over PostgreSQL, and marginal improvements against the commercial system. This demonstrates Bao's ability to adapt to changing data (Stack), and to a significant schema change (Corp), where Bao achieves a 50\% and 40\% reduction in both cost and workload latency (respectively).

\sparagraph{Tail latency analysis} The previous two experiments demonstrate Bao's ability to reduce the cost and latency of an entire workload. Since practitioners are often interested in tail latency (e.g., for an analytics dashboard that does not load until a set of queries is complete, the tail performance essentially determines the performance for the entire dashboard), here we will examine the distribution of query latencies within the IMDb workload on each VM type. Figure~\ref{fig:curves} shows median, 95\%, 99\%, and 99.5\% latencies for each VM type (column) for both PostgreSQL (top row) and the commercial system (bottom row).

For each VM type, \emph{Bao drastically decreases tail latencies when compared to the PostgreSQL optimizer.} On an N1-8 instance, 99\% latency fell from 130 seconds with the PostgreSQL optimizer to under 20 seconds with Bao. This suggests that most of the cost and performance gains from Bao come from reductions at the tail. Compared with the commercial system, Bao always reduces tail latency, although it is only significantly reduced on the smaller VM types. This suggests that the developers of the commercial system may not have invested as much time in reducing tail latencies on less powerful machines: a task that Bao can perform automatically and without any developer intervention.

\begin{figure*}
  \begin{subfigure}{0.24\textwidth}
    \includegraphics[width=\textwidth]{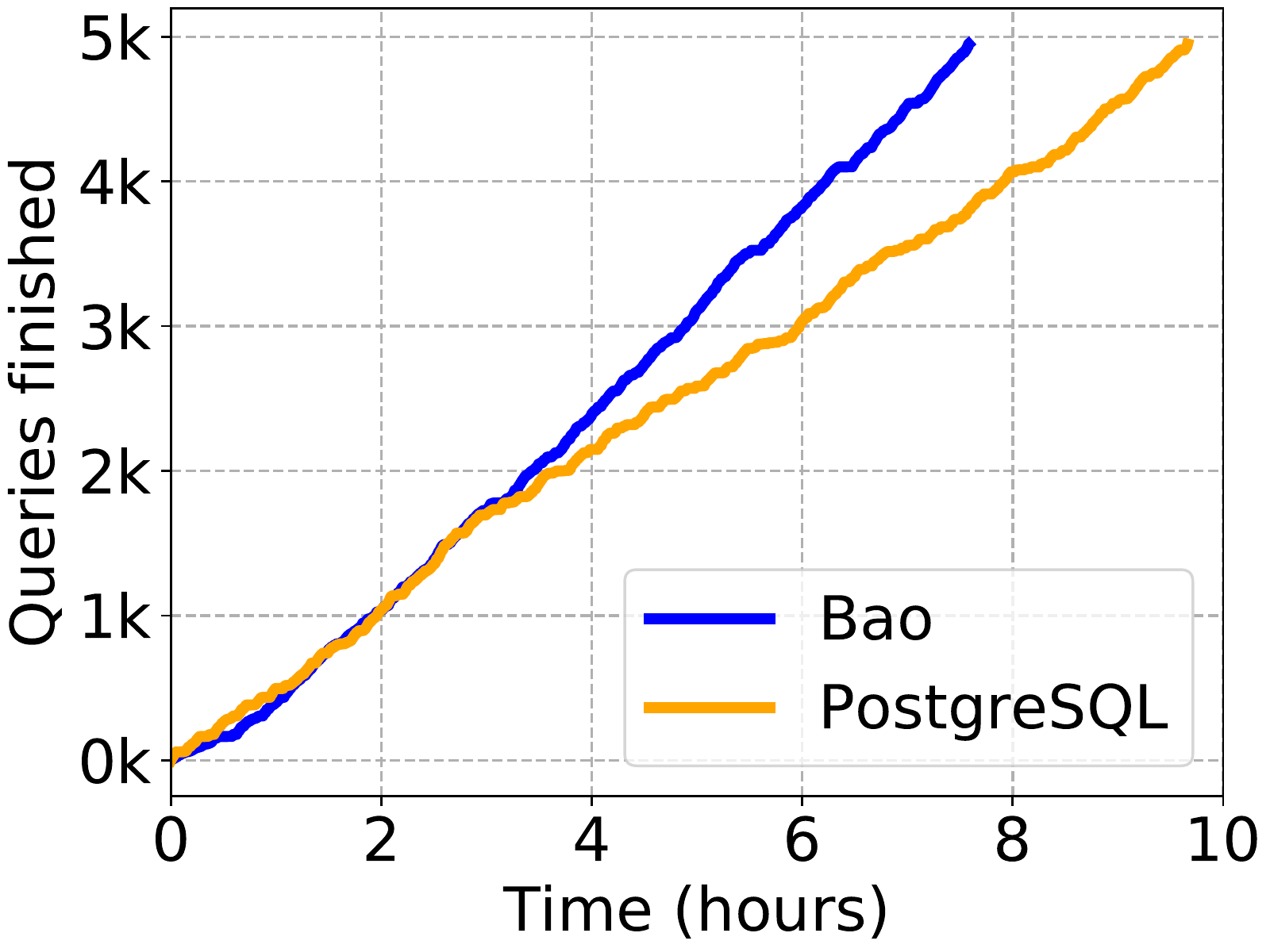}
    \caption{VM type N1-2}
    \label{fig:perf_n1-2}
  \end{subfigure}
  \begin{subfigure}{0.24\textwidth}
    \includegraphics[width=\textwidth]{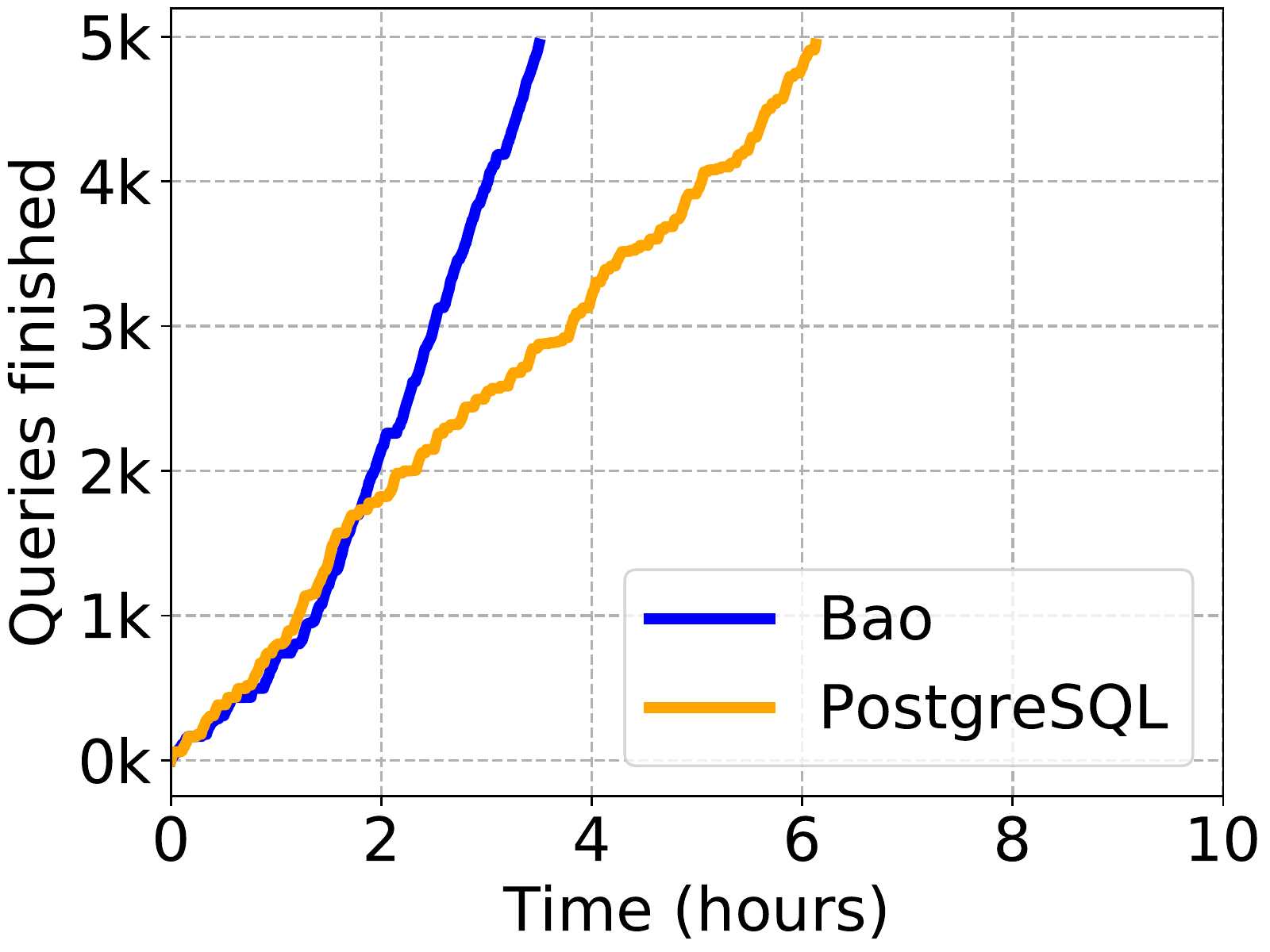}
    \caption{VM type N1-4}
    \label{fig:perf_n1-4}
  \end{subfigure} 
  \begin{subfigure}{0.24\textwidth}
    \includegraphics[width=\textwidth]{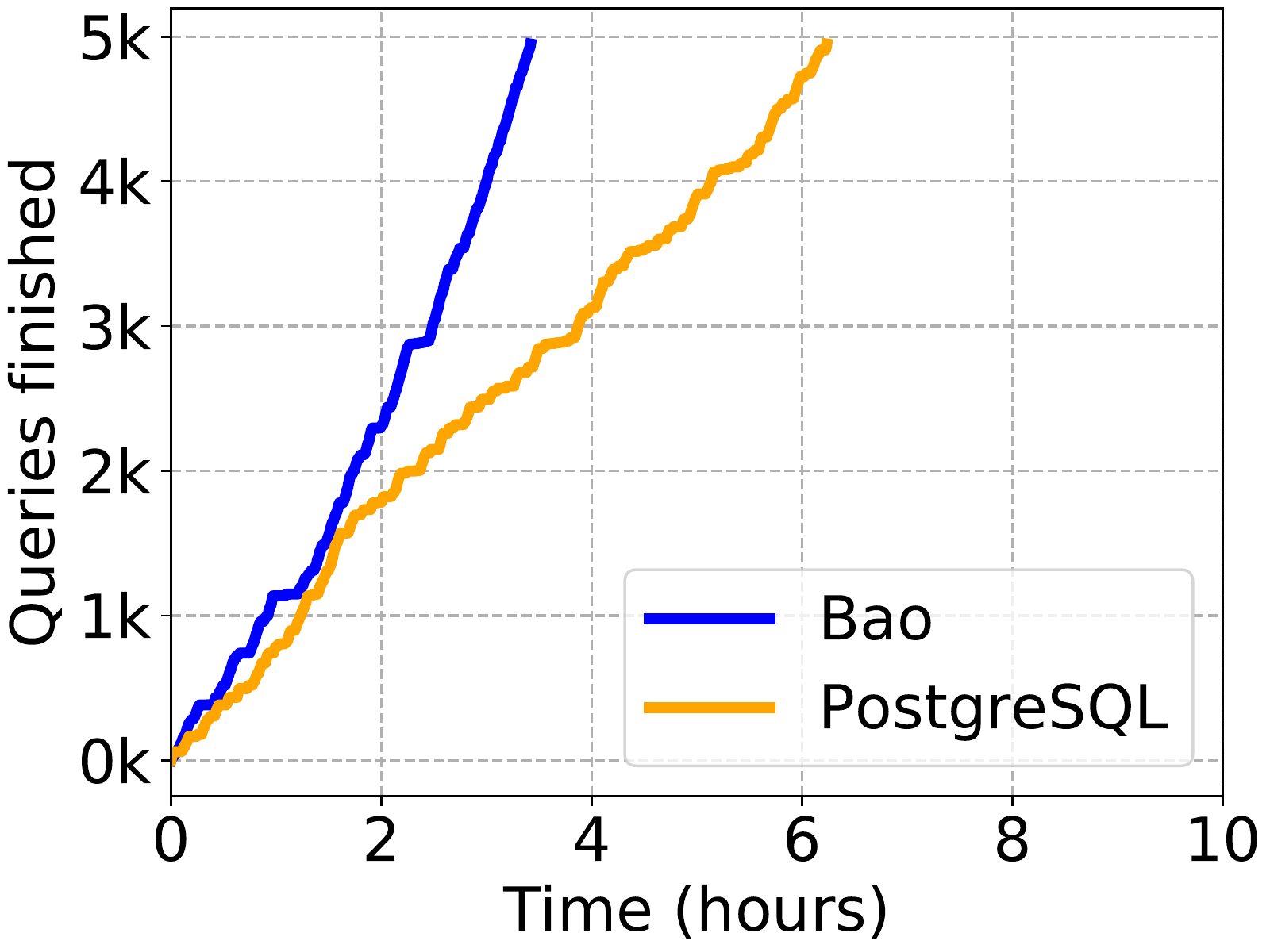}
    \caption{VM type N1-8}
    \label{fig:perf_n1-8}
  \end{subfigure} 
  \begin{subfigure}{0.24\textwidth}
    \includegraphics[width=\textwidth]{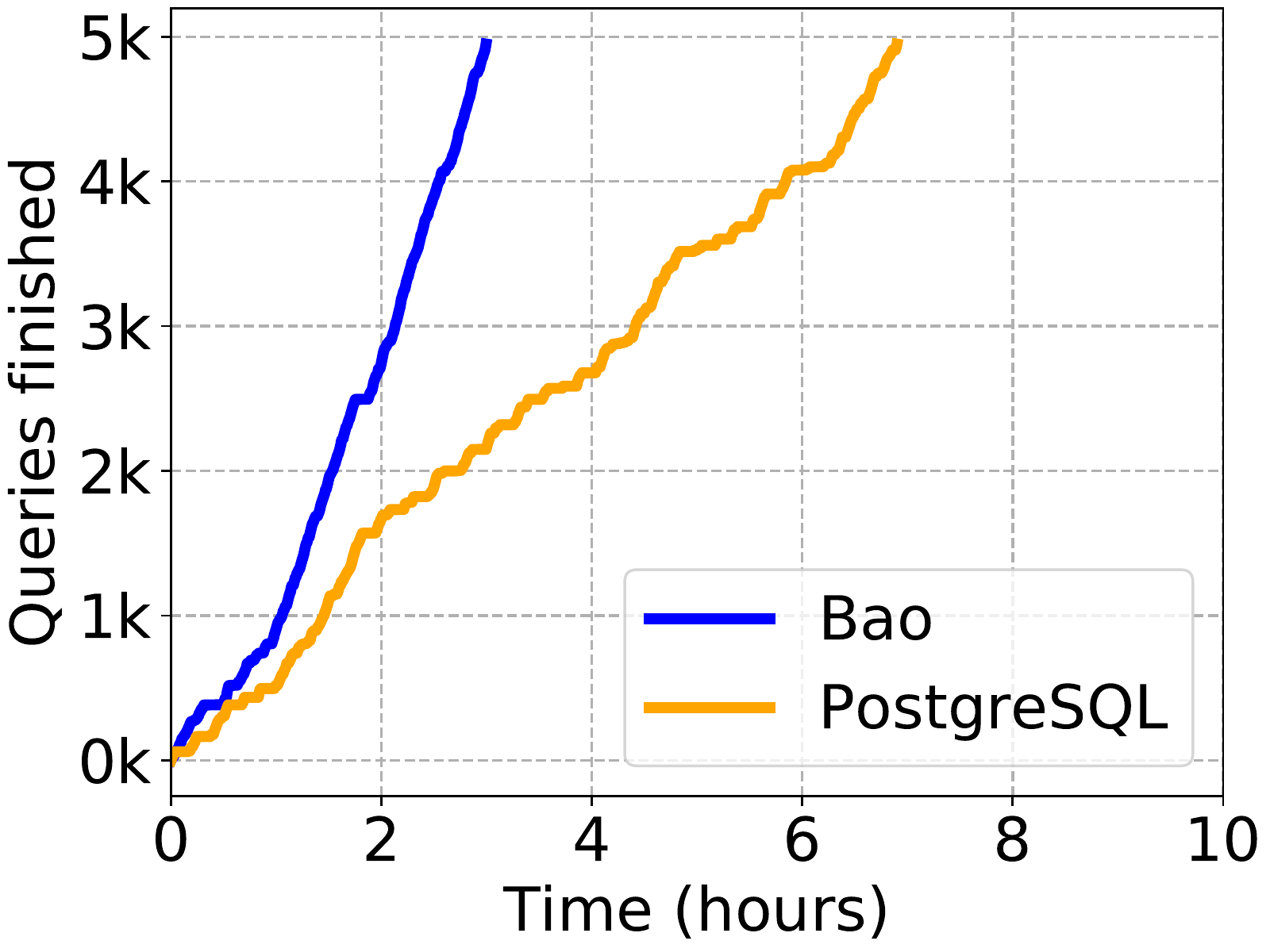}
    \caption{VM type N1-16}
    \label{fig:perf_n1-16}
  \end{subfigure} 
  \caption{Number of IMDb queries processed over time for Bao and the PostgreSQL optimizer on the PostgreSQL engine. The IMDb workload contains 5000 unique queries which vary over time.}
  \label{fig:pref_curves}
\end{figure*}

\sparagraph{Training time and convergence} A major concern with any application of reinforcement learning is convergence time. Figure~\ref{fig:pref_curves} shows time vs. queries completed plots (performance curves) for each VM type while executing the IMDb workload. In all cases, Bao, from a cold start, matches the performance of PostgreSQL within an hour, and exceeds the performance of PostgreSQL within two hours. Plots for the Stack and Corp datasets are similar. Plots comparing Bao against the commercial system are also similar, with slightly longer convergence times: 90 minutes at the latest to match the performance of the commercial optimizer, and 3 hours to exceed the performance of the commercial optimizer.

Note that the IMDb workload is dynamic, and that Bao maintains and adapts to changes in the query workload. This is visible in Figure~\ref{fig:pref_curves}: Bao's performance curve remains straight after a short initial period, indicating that shifts in the query workload did not produce a significant change in query performance.

\begin{figure*}
  \centering
  \begin{subfigure}{0.32\textwidth}
    \includegraphics[width=\textwidth]{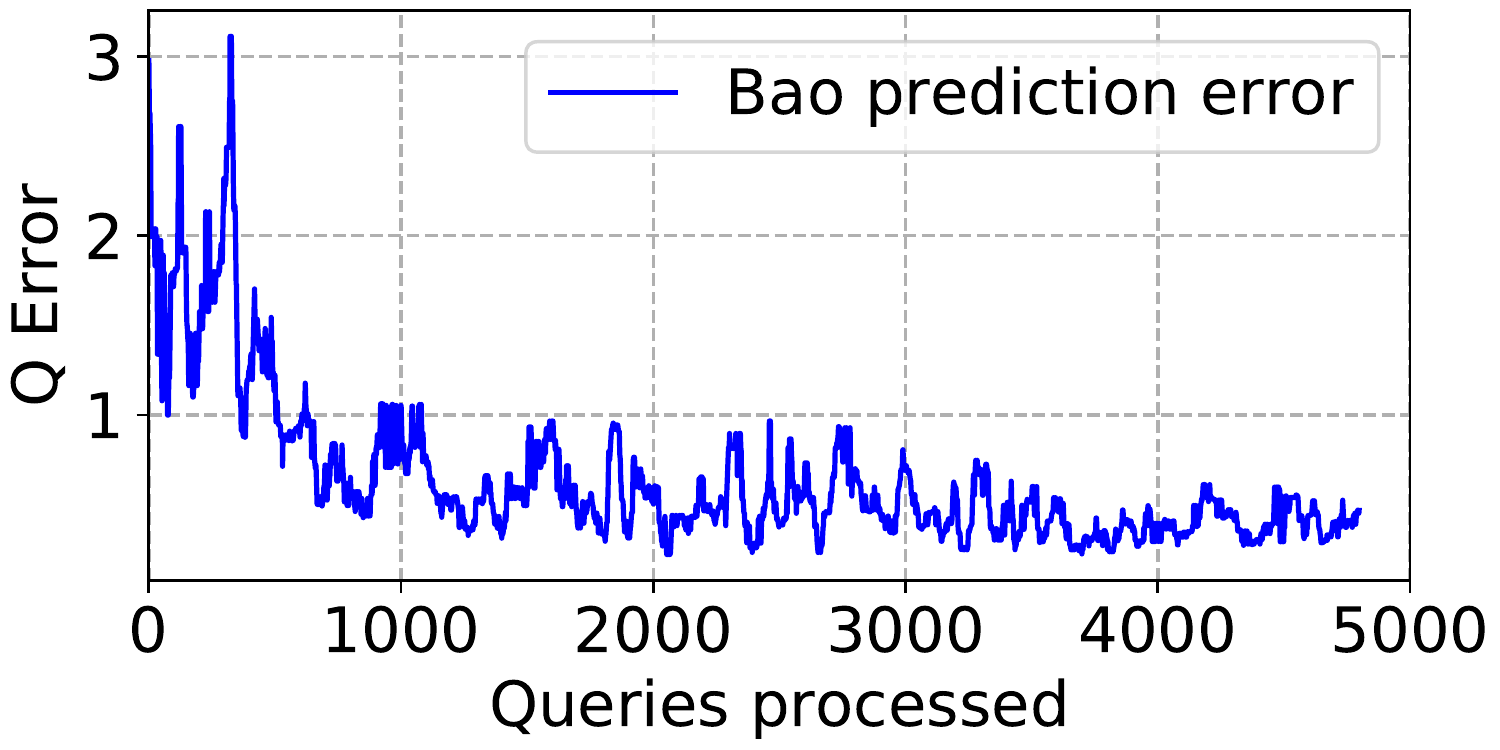}
    \caption{Median Q-Error (0 is a perfect prediction) of Bao's predictive model vs. the number of queries processed. IMDb workload on N1-16 VM using PostgreSQL engine.}
    \label{fig:accuracy}
  \end{subfigure}
  \begin{subfigure}{0.32\textwidth}
    \includegraphics[width=\textwidth]{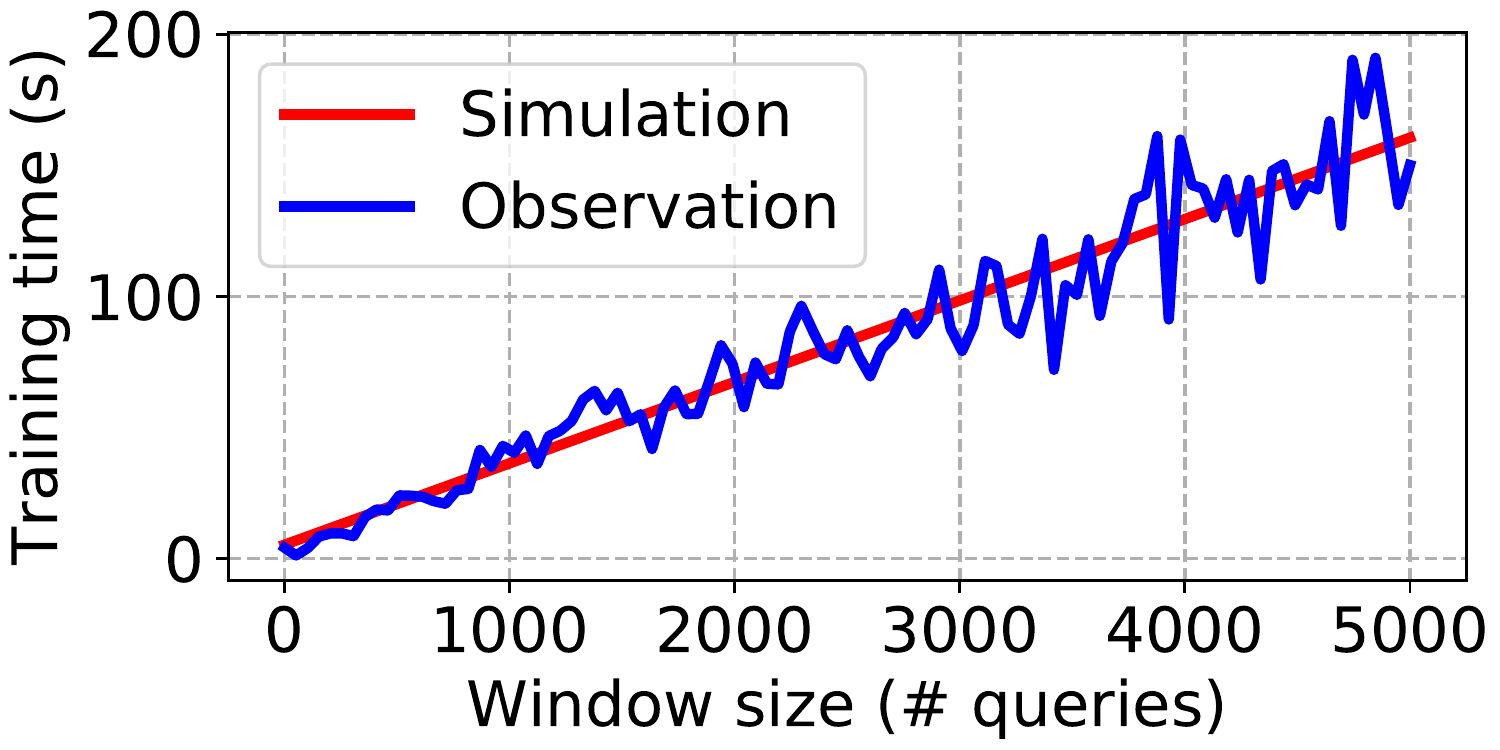}
    \caption{Simulated and observed time to train Bao's performance prediction model (GPU) based on the sliding window size (number of queries used during each training iteration).}
    \label{fig:training}
  \end{subfigure}
  \begin{subfigure}{0.32\textwidth}
    \includegraphics[width=\textwidth]{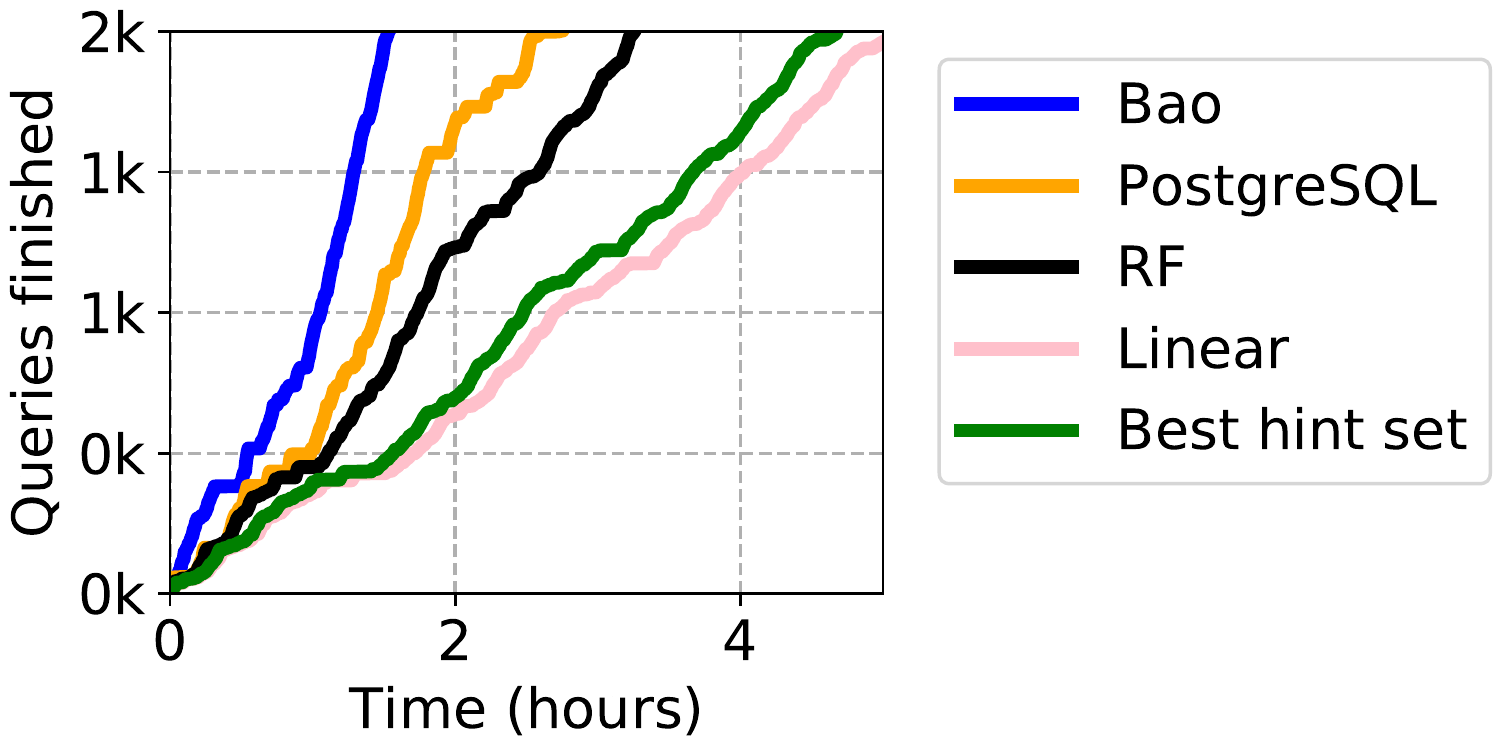}
    \caption{Random forest (RF) and linear models (Linear) used as Bao's model. ``Best hint set'' is the single best hint set. IMDb, N1-16 VM, on PostgreSQL.}
    \label{fig:non_nn}
  \end{subfigure}
  \caption{}
\end{figure*}

\sparagraph{Predictive model accuracy} The core of Bao's bandit algorithm is a predictive model, which Bao uses to select hint sets for each incoming query. As Bao makes more decisions, Bao's experience grows, allowing Bao to train a more accurate model. Figure~\ref{fig:accuracy} shows the accuracy of Bao's predictive model after processing each query in the IMDb workload on an N1-16 machine. Following~\cite{qppnet, li_robust}, we use Q-Error instead of relative error~\cite{no_percent_error}. Given a prediction $x$ and a true value $y$, the Q-Error is defined as:
\begin{equation*}
  QError(x, y) = \max\left(\frac{x}{y}, \frac{y}{x}\right) - 1
\end{equation*}
Q-Error can be interpreted as a symmetric version of relative error. For example, a Q-Error of 0.5 means the estimator under or over estimated by 50\%.

As shown in Figure~\ref{fig:accuracy}, Bao's predictive model begins with comparatively poor accuracy, with a peak misprediction of 300\%. Despite this inaccuracy, Bao is still able to choose plans that are not catastrophic (as indicated by Figure~\ref{fig:perf_n1-16}). Bao's accuracy continues to improve as Bao gains more experience. We note that this does not indicate that Bao could be used as generic query performance prediction technique: here, Bao's predictive model is only evaluated on query plans produced by one optimizer, and thus most of the plans produced may not be representative of what other query optimizer's produce.

\sparagraph{Required GPU time} Bao's tree convolution neural network is trained on a GPU. Because attaching a GPU to a VM incurs a fee, Bao only keeps a GPU attached for the time required to train the predictive model (see Section~\ref{sec:train_loop}). Figure~\ref{fig:training} shows how long this training time takes as a function of the window size $k$ (the maximum number of queries Bao uses to train). Theoretically, the training time required should be linear in the window size. The ``Observation'' line shows the average time to train a new predictive model at a given window size. The ``Simulation'' line shows the linear regression line of these data, which shows the training time does indeed follow a linear pattern. Fluctuations may occur for a number of reasons, such as block size transfers to the GPU, noisy neighbors in cloud environments, and the stochastic nature of the Adam optimizer.

Generally speaking, larger window sizes will require longer training, and thus more GPU time, but will provide a more accurate predictive model. While we found a window size of $k=2000$ to work well, practitioners will need to tune this value for their needs and budget (e.g., if one has a dedicated GPU, there may be no reason to limit the window size at all). We note that even when the window size is set to $k=5000$ queries (the maximum value for our workloads with 5000 queries), training time is only around three minutes.

\sparagraph{Do we need a neural network?} Neural networks are a heavy-weight machine learning tool, and should only be applied when simpler techniques fail. Often, simpler techniques can even perform better than deep learning~\cite{msr_paper}. To determine if the specialized tree convolutional neural network used by Bao was justified, we tested random forest (RF) and linear regression (Linear) models as well.\footnote{We performed an extensive grid search to tune the random forest model.} For these techniques, we featurized each query plan tree into a single vector by computing minimums, medians, maximums, averages, and variances of each entry of each tree node's vectorized representation. The resulting performance curve for the first 2000 queries of the IMDb workload is shown in Figure~\ref{fig:non_nn}. Both the random forest and the linear regression model fail to match PostgreSQL's performance.

This provides evidence that a deep learning approach is justified for Bao's predictive model. We note that Bao's neural network is not a standard fully-connected neural network: as explained in~\cite{neo}, tree convolution carries a strong \emph{inductive bias}~\cite{inductive_bias_rl} that matches well with query plans.

\sparagraph{Is one hint set good for all queries?} A simple alternative to Bao might be successive elimination bandit algorithms~\cite{suc_elim}, which seek to find the single besthint set regardless of a specific query. We evaluated each hint set on the entire IMDb workload. In Figure~\ref{fig:non_nn}, we plot this single best hint set (which is disabling loop joins) as ``Best hint set''. This single hint set, while better than all the others, performs significantly worse than the PostgreSQL optimizer. Since, unsurprisingly, no single hint set is good enough to outperform PostgreSQL (otherwise, PostgreSQL would likely use it as a default), we conclude that successive elimination bandit algorithms are not suitable for this application.

This also provides evidence that Bao is learning a non-trivial strategy when selecting hint sets, as if Bao were only selecting one hint set, Bao's performance could not possibly be better than ``Best hint set.'' To further test this hypothesis, we evaluated the number of distinct hint sets chosen more than 100 times for each dataset on a N1-4 machine on the PostgreSQL engine. For IMDb, $\frac{35}{48}$ hint sets were chosen over 100 times, indicating a high amount of diversity to Bao's strategy. For Stack, this value was $\frac{37}{48}$, and for Corp this value was $\frac{15}{48}$. The less diverse strategy learned by Bao in the Corp case could be due to similarities between queries issued by analysts. We leave a full analysis of Bao's learned strategies, and their diversity, to future work.


\begin{figure}
  \centering
  \includegraphics[width=0.35\textwidth]{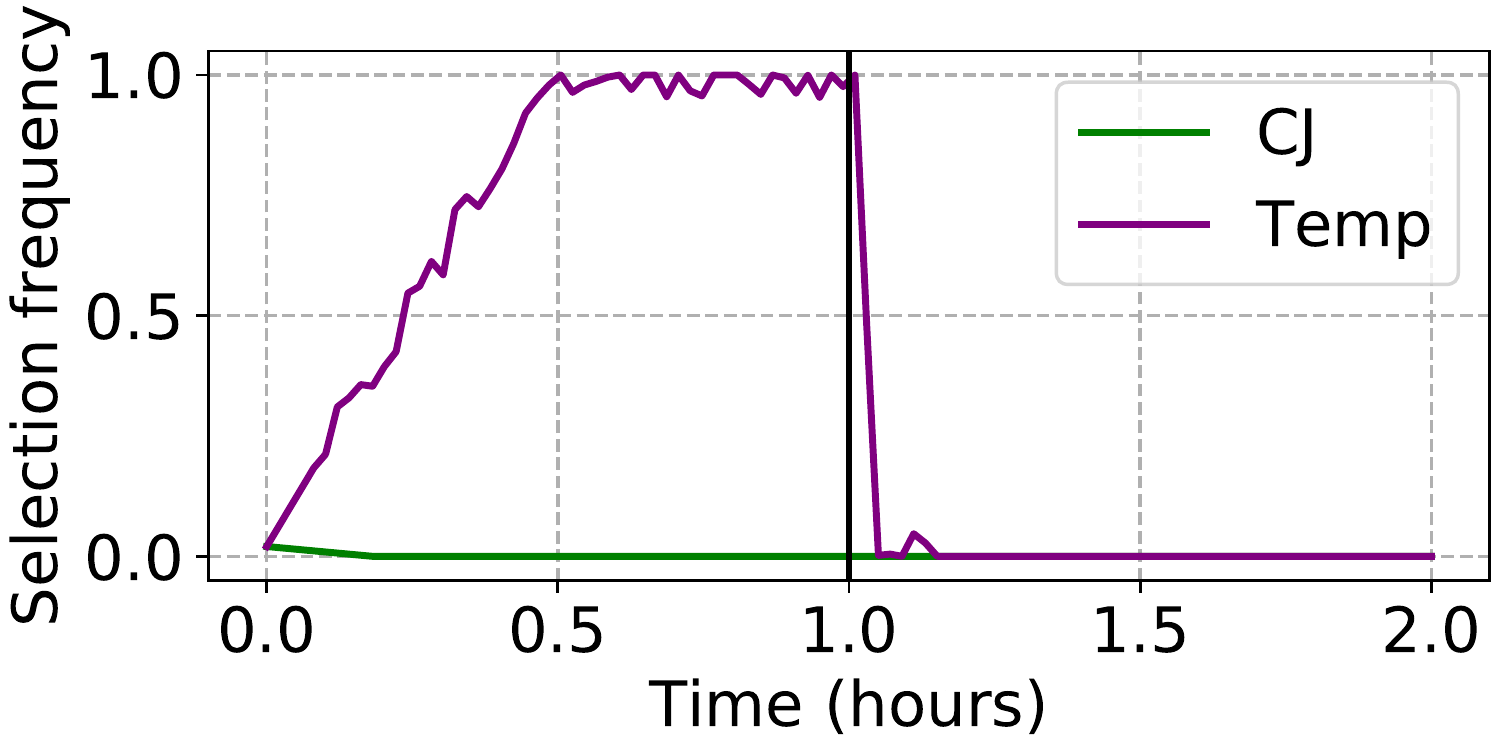}
  \caption{Selection frequency of optimizers over time when Bao is given a specific additional hitn set. ``CJ'' is a hint set that always induces plans with cross joins. ``Temp'' is hint set that induces the optimal plan until one hour has elapsed, then induces plans with cross joins.}
  \label{fig:bad_opt}
\end{figure}

\sparagraph{What if a hint set performs poorly?} Here,  we test Bao's resiliency to hint sets that induce unreasonable query plans by artificially introducing a poor-performing hint set. Figure~\ref{fig:bad_opt} shows the selection frequency (how often Bao chooses a particular hint set over a sliding window of 100 queries) over a two-hour period executing the IMDb workload. The first hint set, ``CJ'', induces query plans with cross joins that are 100x worse than the optimal. Bao selects the ``CJ'' hint set once (corresponding to an initial selection frequency of $\frac{1}{49}$), and then never selects the ``CJ'' hint set again for the duration of the experiment. Bao is able to learn to avoid ``CJ'' so efficiently (i.e., with extremely few samples) because Bao makes predictions based on query plans: once Bao observes a plan full of cross joins, Bao immediately learns to avoid plans with cross joins.

It is possible that a hint set's behavior changes over time, or that a change in workload or data may cause a hint set to become unreasonable. For example, if new data was added to the database that resulted in a particular intermediary becoming too large to fit in a hash table, a hint set inducing plans that use exclusively hash joins may no longer be a good choice.

To test Bao's resiliency to this scenario, we introduce the ``Temp'' hint set, which produces the optimal query plan (precomputed ahead of time) until 1 hour has elapsed, at which point the ``Temp'' hint set induces query plans with cross joins (the same plans as ``CJ''). Figure~\ref{fig:bad_opt} shows that, in the first hour, Bao learns to use ``Temp'' almost exclusively (at a rate over 95\%), but, once the behavior of ``Temp'' changes, Bao quickly stops using ``Temp'' entirely.

\sparagraph{Optimization time} Another concern with applying machine learning to query optimization is inference time. Surprisingly, some reinforcement learning approachs~\cite{sanjay_wat, rejoin} actually decrease optimization time. Here, we evaluate the optimization overhead of Bao. Across all workloads, the PostgreSQL optimizer had a maximum planning time of 140ms. The commercial system had a maximum planning time of 165ms. Bao had higher maximum planning times than both other systems, with a maximum planning time of 210ms. Bao's increased planning time is due to two factors:
\begin{enumerate}
\item{Hint sets: a query plan must be constructed for each hint set. While each hint set can be ran in parallel, this accounts for approximately 80\% of Bao's planning time (168ms).}
\item{Neural network inference: after each hint set produces a query plan, Bao must run each one through a tree convolutional neural network. These query plans can be processed in a batch, again exploiting parallelism. This accounts for the other 20\% of Bao's planning time (42ms.)}
\end{enumerate}

Since analytic queries generally run for many seconds or minutes, a 210ms optimization time may be acceptable in some applications. Further optimizations of the query optimizer, or optimizations of the neural network inference code (written in Python for our prototype), may reduce optimization time.  Applications requiring faster planning time may wish to consider other options~\cite{quickpick}.

\begin{table}
\centering
\begin{tabularx}{0.5\textwidth}{lXXX}
\toprule
  & {\bf Trad} & {\bf Neo} & {\bf Bao} \\
\midrule
{\bf Needs cardinality estimation} & Yes & No             & Yes \\
{\bf Needs cost model}             & Yes & No             & Yes \\
{\bf Needs pretraining}            & No  & Yes            & No  \\
\midrule
{\bf Handles schema changes}          & Yes & No      & Yes \\
{\bf Handles data changes}            & Yes & No      & Yes \\
{\bf Handles workload changes}        & Yes & Slowly  & Yes \\
{\bf Accounts for cache}              & Maybe  & No      & Yes \\
\midrule
{\bf Approx. convergence time}     & 0 hrs & 24 hrs       & 1 hr \\
\bottomrule
\end{tabularx}
\caption{Requirements and feature comparison of a traditional cost-based optimizer (Trad), the fully-learned optimizer Neo~\cite{neo}, and Bao.}
\label{tab:compare}
\end{table}

\begin{figure}
  \centering
  \begin{subfigure}{0.23\textwidth}
    \includegraphics[width=\textwidth]{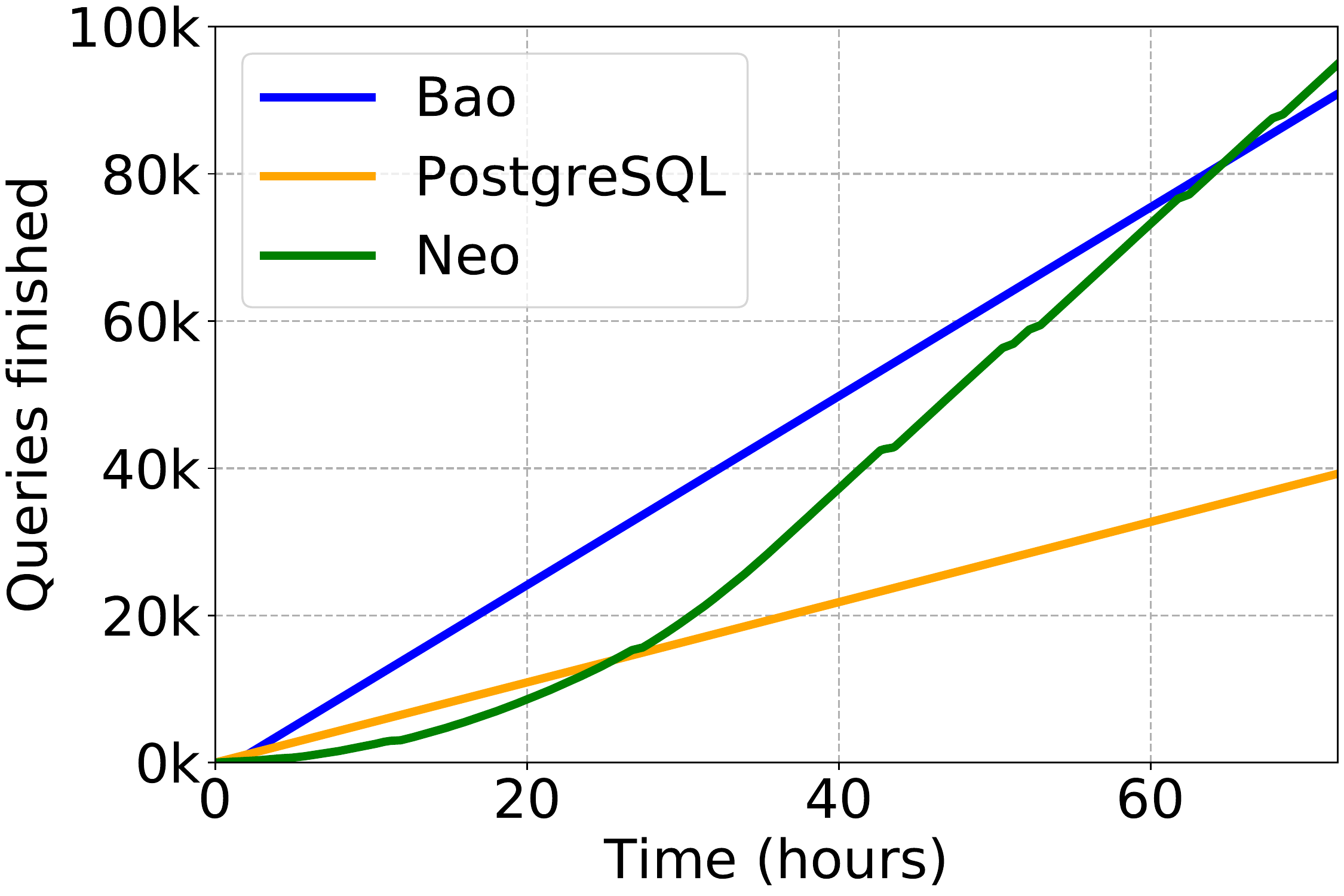}
    \caption{Stable query workload}
    \label{fig:neo_stable}
\end{subfigure} 
\begin{subfigure}{0.23\textwidth}
    \includegraphics[width=\textwidth]{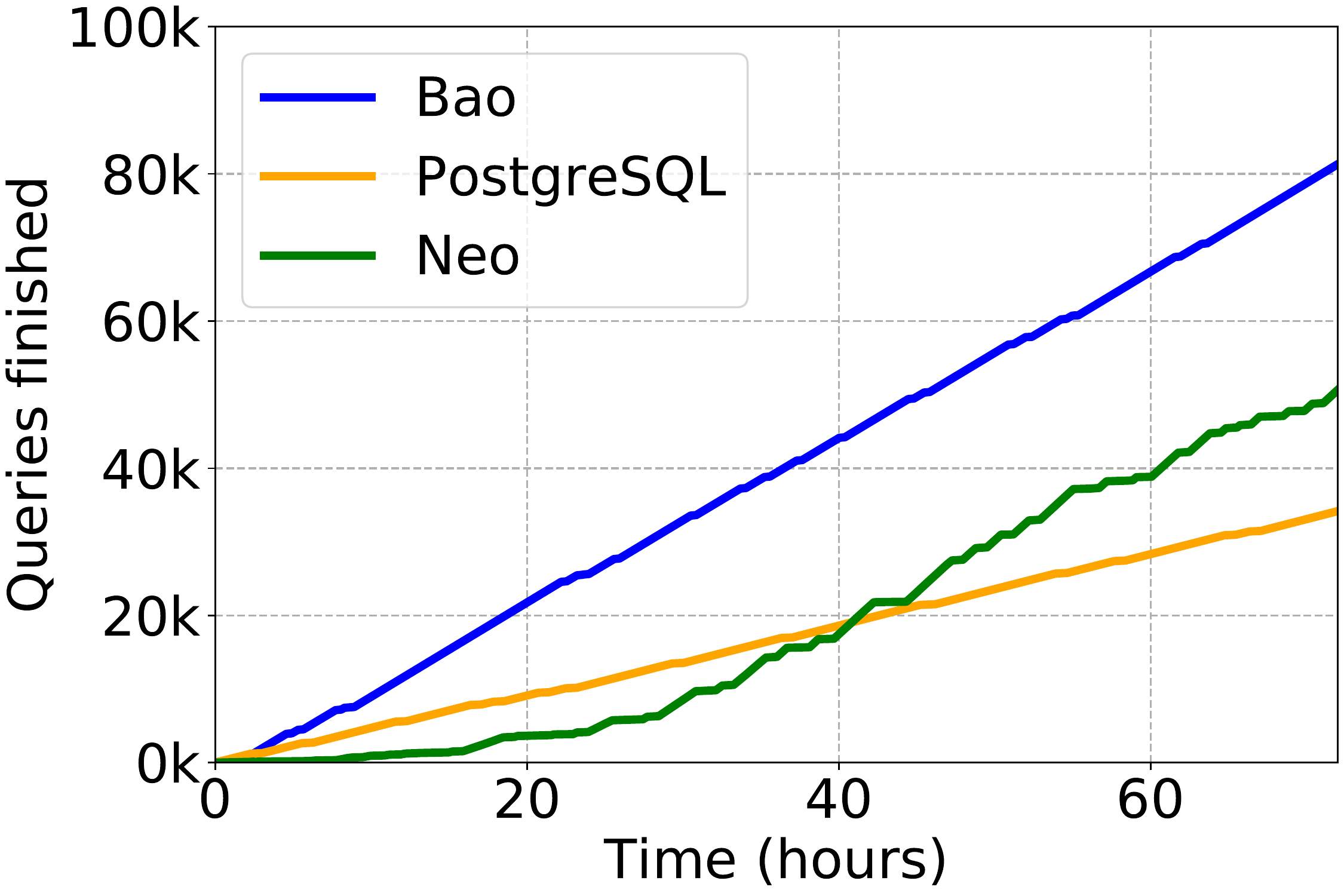}
    \caption{Dynamic query workloads}
    \label{fig:neo_unstable}
  \end{subfigure} 
\caption{Comparison of number of queries finished over time for Bao, Neo, and PostgreSQL for a stable query workload (left) and a dynamic query workload (right).}
\label{fig:neo}
\end{figure}

\begin{figure}
  \includegraphics[width=0.49\textwidth]{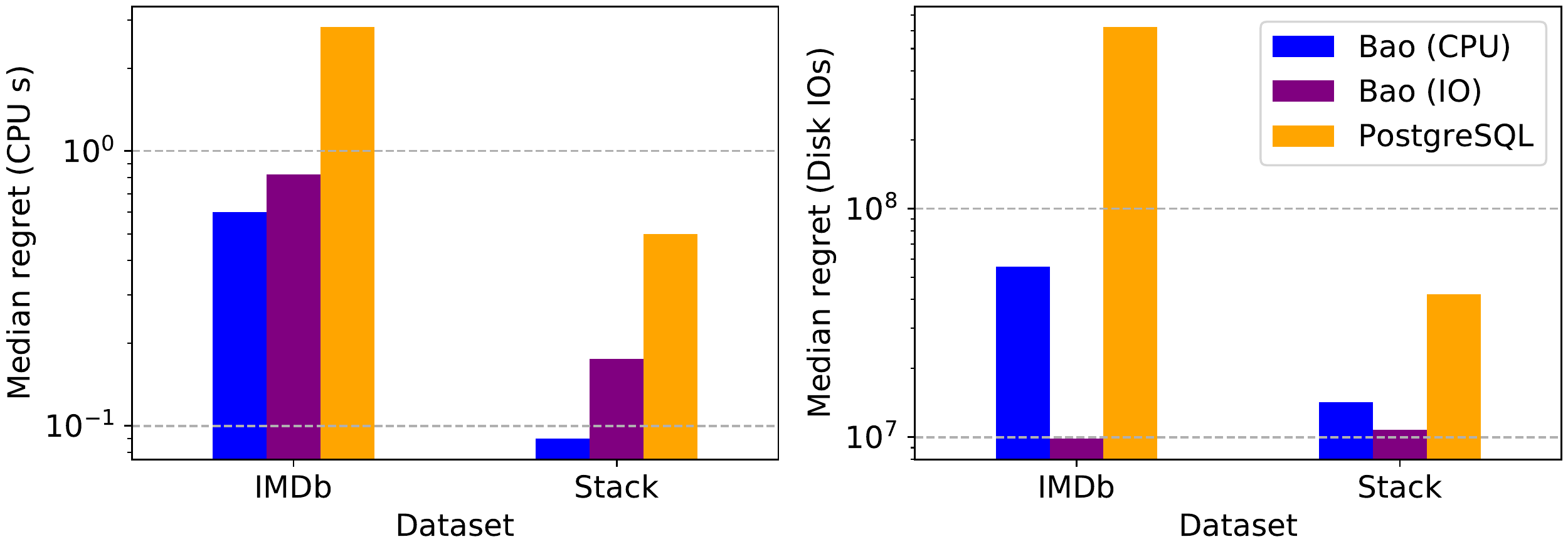}
  \caption{\hl{Comparison of the median regret (difference between the outcome of the selected action and the ideal hint set) when Bao is optimizing for CPU time or for physical disk I/Os. The two Bao models and PostgreSQL are plotted in terms of CPU time (left) and physical disk I/Os (right). Executed on the PostgreSQL engine.}}
  \label{fig:median_regret}
\end{figure}

\begin{figure*}[t]
  \includegraphics[width=\textwidth]{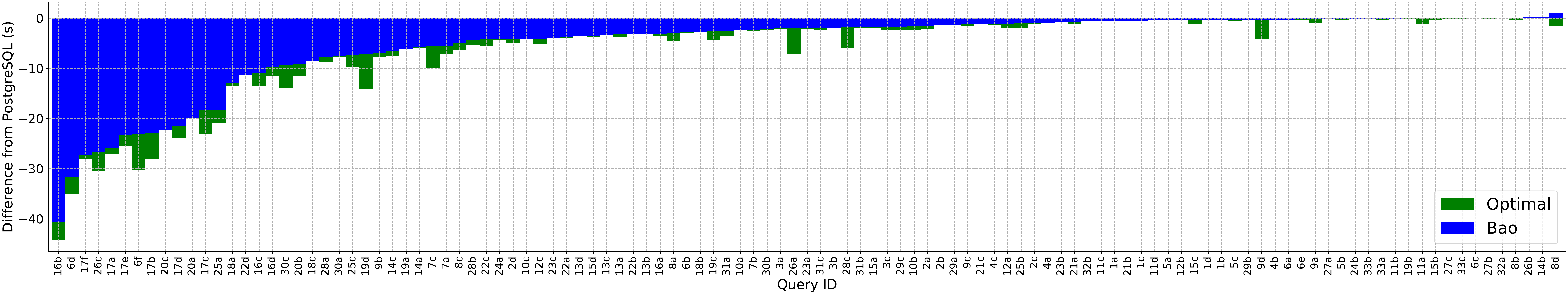}
  \caption{Absolute difference in query latency between Bao's selected plan and PostgreSQL's selected plan for the subset of the IMDb queries from the Join Order Benchmark~\cite{howgood} (lower is better).}
  \label{fig:by_query}
\end{figure*}

\sparagraph{Comparison with Neo} Neo~\cite{neo} is an end-to-end query optimizer based on deep reinforcement learning. Like Bao, Neo uses tree convolution, but unlike Bao, Neo does not select hint sets for specific queries, but instead fully builds query execution plans on its own. A qualitative comparison of Neo, Bao, and traditional query optimizers is shown in Table~\ref{tab:compare}. While Neo avoids the dependence on a cardinality estimator and a cost model (outside of Neo's bootstrapping phase), Neo is unable to handle schema changes or changes in the underlying data (to handle these scenarios, Neo requires retraining). Additionally, because Neo is learning a policy to construct query plan trees themselves (a more complex task than choosing hint sets), Neo requires substantially longer training time to match the performance of traditional query optimizers (i.e., 24 hours instead of 1 hour). While neither the PostgreSQL nor the ComSys optimizer took cache state into account, implementing cache awareness in a traditional cost-based optimizer is theoretically possible, although likely difficult.

In Figure~\ref{fig:neo}, we present a quantitative comparison of Bao and Neo. Each plot shows the performance curves for the IMDb workload repeated 20 times on an N1-16 machine with a cutoff of 72 hours. However, in Figure~\ref{fig:neo_stable}, we modify the IMDb workload so that each query is chosen uniformly at random (i.e., the workload is no longer dynamic). With a stable workload, Neo is able to overtake PostgreSQL after 24 hours, and Bao after 65 hours. This is because Neo has many more degrees of freedom than Bao: Neo can use any logically correct query plan for any query, whereas Bao is limited to a small number of options. However, these degrees of freedom come with a cost, as Neo takes significantly longer to converge. When the workload, schema, and data are all stable, and a suitable amount of training time is available, the plans learned by Neo are superior to the plans selected by Bao, and Neo will perform better over a long time horizon.

In Figure~\ref{fig:neo_unstable}, we use a dynamic workload instead of a static workload. In this case, Neo's convergence is significantly hampered: Neo requires much more time to learn a policy robust to the changing workload and overtake PostgreSQL (42 hours). With a dynamic workload, Neo is unable to overtake Bao. This showcases Bao's ability to adapt to changes better than previous learned approaches.

\begin{figure}
\begin{subfigure}{0.49\textwidth}
    \includegraphics[width=\textwidth]{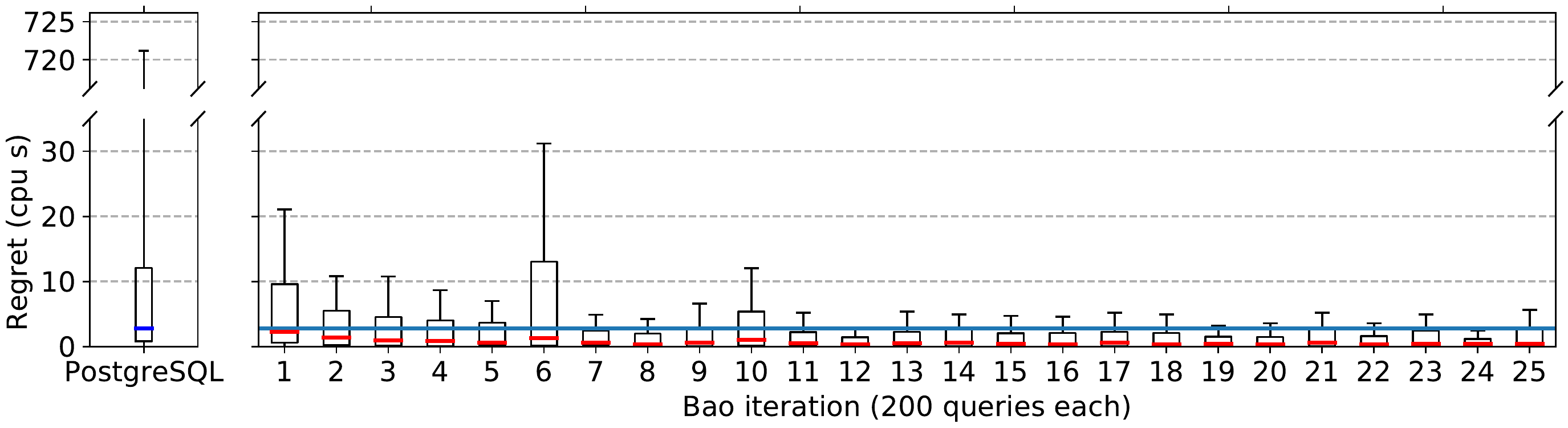}
    \caption{CPU time regret, IMDb}
    \label{fig:cpu_job}
\end{subfigure} 
\begin{subfigure}{0.49\textwidth}
    \includegraphics[width=\textwidth]{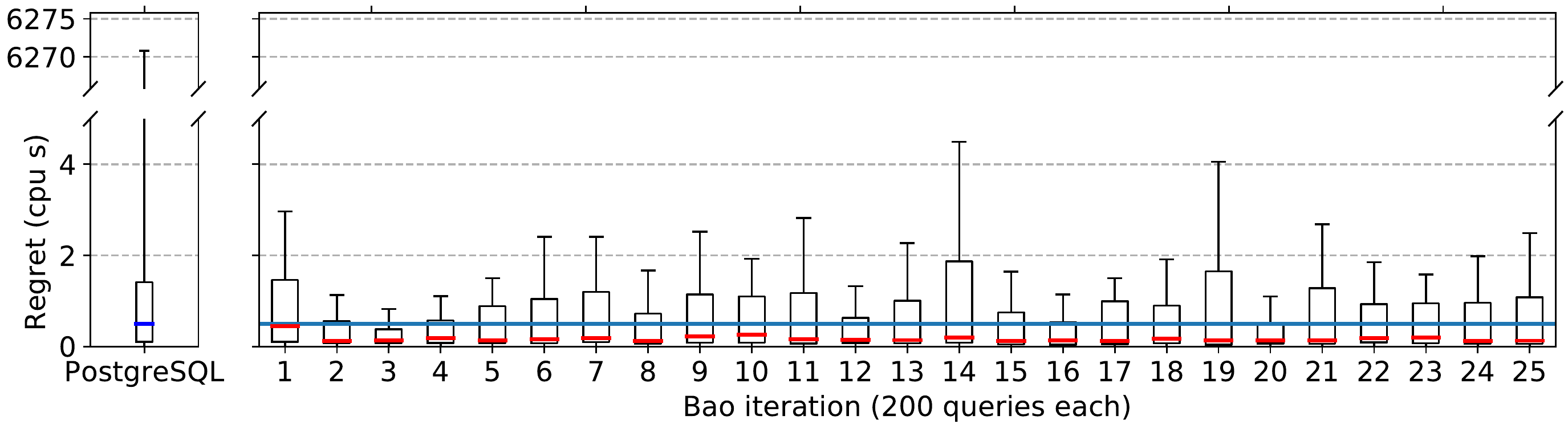}
    \caption{CPU time regret, StackOverflow}
    \label{fig:cpu_so}
  \end{subfigure} 
\caption{CPU time regret (difference between the outcome from the optimal hint set and the selected hint set), when Bao is optimizing for CPU time. Bao (right) is compared against the PostgreSQL optimizer (left) on the PostgreSQL engine over multiple training iterations of 50 queries each. The blue line marks the median regret of the PostgreSQL optimizer. Note the cut axes. Whiskers show the 98\% percentile.}
\label{fig:cpu_regret}
\end{figure}

\begin{figure}
\begin{subfigure}{0.49\textwidth}
    \includegraphics[width=\textwidth]{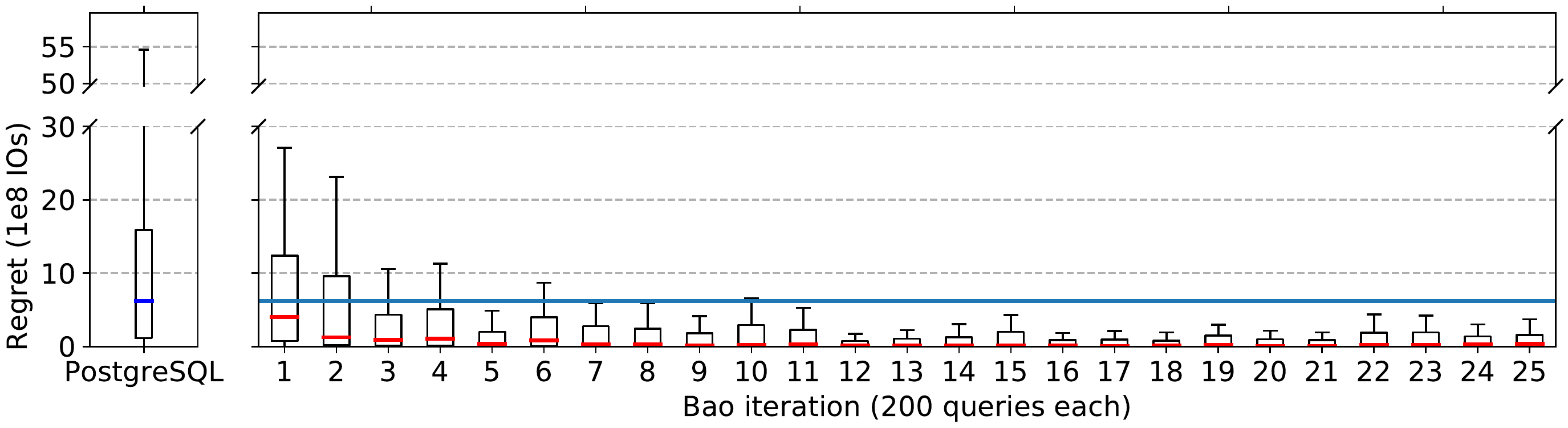}
    \caption{Physical I/O regret, IMDb}
    \label{fig:io_job}
\end{subfigure} 
\begin{subfigure}{0.49\textwidth}
    \includegraphics[width=\textwidth]{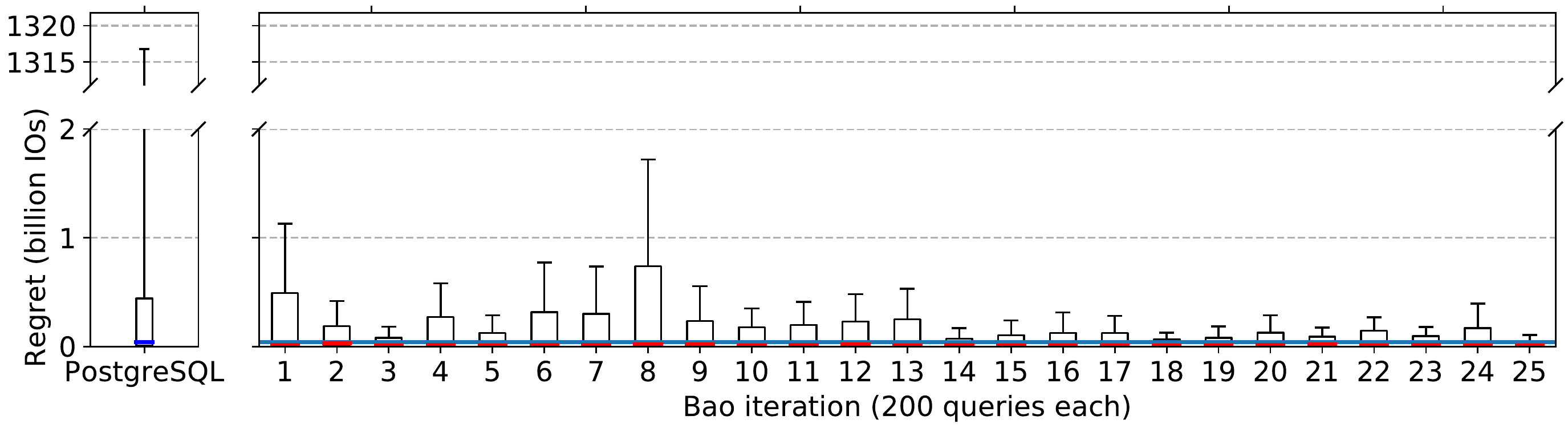}
    \caption{Physical I/O regret, StackOverflow}
    \label{fig:io_so}
  \end{subfigure} 
\caption{Physical I/O request regret (difference between the number of physical I/O requests made by the optimal hint set and the selected hint set), when Bao is optimizing for physical I/Os. Bao (right) is compared against the PostgreSQL optimizer (left) on the PostgreSQL engine over multiple training iterations of 50 queries each. The blue line marks the median regret of the PostgreSQL optimizer. Note the cut axes. Whiskers show the 98\% percentile.}
\label{fig:io_regret}
\end{figure}

\subsection{Optimality}
\label{sec:cold}

Here, we evaluate Bao's \emph{regret}, the difference in performance relative to the optimal hint set for each query. The optimal hint set for each query was computed by exhaustively executing all query plans. In order for this to be practical, each query is executed with a cold cache\footnote{Since different query plans may behave differently with different cache states, a cold cache is required to keep the computation feasible.} on a cluster of GCP~\cite{url-google} nodes. 

\sparagraph{Customizable optimization goals} Here, we test Bao's ability to optimize for different metrics. Figure~\ref{fig:median_regret} shows the median regret observed over all IMDb and Stack queries. \hl{We train two different Bao models, one which optimizes for CPU time (``Bao (CPU)''), and one which optimizers for disk IOs (``Bao (IO)''). Figure~\ref{fig:median_regret} shows the median regret of these two Bao models and the PostgreSQL optimizer in terms of CPU time (left) and disk IOs (right). Unsurprisingly, Bao achieves a lower median CPU time regret when trained to minimize CPU time, and Bao achieves a lower median disk IO regret when trained to minimize disk IOs.} Incidentally, for both metrics and both datasets, Bao achieves a significantly lower median regret than PostgreSQL, \emph{\hl{regardless of which metric Bao is trained on.}}. The ability to customize Bao's performance goals could be helpful for cloud providers with complex, multi-tenant resource management needs.

\sparagraph{Regret over time \& tails} Figure~\ref{fig:cpu_regret}~and~\ref{fig:io_regret} shows the distribution of regret for both PostgreSQL (left) and Bao over each iteration (right). Note both the cut axes and that the whiskers show the 98\% percentile. For both metrics and datasets, Bao is able to achieve significantly better tail regret \emph{from the first iteration} (after training). For example, when optimizing CPU time, the PostgreSQL optimizer picks several query plans requiring over 720 CPU seconds, whereas Bao never chooses a plan requiring more than 30 CPU seconds. The improvement in the tail of regret is similar for both metrics and datasets.

Figure~\ref{fig:cpu_regret}~and~\ref{fig:io_regret} also show that Bao quickly matches or beats the median regret of the PostgreSQL optimizer (in the case of physical I/Os for Stack, both Bao and PostgreSQL achieve median regrets near zero). Median regret may be more important than tail regret in single-tenant settings. This demonstrates that Bao, in terms of both the median and the tail regret, is capable of achieving lower regret than the PostgreSQL optimizer.

\sparagraph{Query regression analysis} Finally, we analyze the per-query performance of Bao. Figure~\ref{fig:by_query} shows, for both Bao and the optimal hint set, the absolute performance improvement (negative) or regression (positive) for each of the Join Order Benchmark (JOB)~\cite{howgood} queries (a subset of our IMDb workload). For this experiment, we train Bao by executing the entire IMDb workload with the JOB queries removed, and then executed each JOB query without updating Bao's predictive model (in other words, our IMDb workload without the JOB query was the training set, and the JOB queries were the test set). There was no overlap in terms of predicates or join graphs. Of the 113 JOB queries, Bao only incurs regressions on three, and these regressions are all under 3 seconds. Ten queries see performance improvements of over 20 seconds. While Bao (blue) does not always choose the optimal hint set (green), Bao does come close on almost every query. Interestingly, for every query, one hint set was always better than the plan produced by PostgreSQL, suggesting that a perfect Bao model could achieve zero regressions.


\vspace{2mm}
\section{Conclusion and future work}
\label{sec:conclusion}

This work introduced Bao, a bandit optimizer which steers a query optimizer using reinforcement learning. Bao is capable of matching the performance of open source and commercial optimizers with as little as one hour of training time. Bao uses a combination of Thompson sampling and tree convolutional neural networks to select query-specific optimizer hints. We have demonstrated that Bao can reduce median and tail latencies, even in the presence of dynamic workloads, data, and schema.

In the future, we plan to more fully investigate integrating Bao into cloud systems. Specifically, we plan to test if Bao can improve resource utilization in multi-tenant environments where disk, RAM, and CPU time are scarce resources. We additionally plan to investigate if Bao's predictive model can be used as a cost model in a traditional database optimizer, enabling more traditional optimization techniques to take advantage of machine learning.


\balance

\bibliographystyle{abbrv}
\bibliography{ryan-cites-long}

\end{document}